\documentstyle[12pt,a4,epsf]{article}
\newcommand{\rcite}[1]{{\cite{#1}}}
\newcommand{\rref}[1]{{(\ref{#1})}}
\newcommand{\tref}[1]{{\ref{#1}}}
\newcommand{\rlabel}[1]{{\label{#1}}}
\newcommand{\rbibitem}[1]{\bibitem{#1}}
\newcommand{\be}{\begin{equation}}
\newcommand{\ee}{\end{equation}}
\newcommand{\ba}{\begin{eqnarray}}
\newcommand{\ea}{\end{eqnarray}}
\newcommand{\dis}{\displaystyle}

\newcommand{\Pids}{\Pi_{\Delta S =2}}
\newcommand{\mathrm}[1]{{\rm #1}}
\newcommand{\tr}{\mathrm{tr}}

\def\theequation{\arabic{section}.\arabic{equation}}

\begin{document}
\begin{titlepage}
\begin{flushright}
{NORDITA-95/11 N,P\\}
{hep-ph/9502363}
\end{flushright}
\vspace{2cm}
\begin{center}
{\large\bf The $B_K$ Parameter in the  $1/N_c$ Expansion}\\
\vfill
{\bf Johan Bijnens$^a$ and Joaqu\'{\i}m Prades$^{a,b}$}\\[0.5cm]
$^a$ NORDITA, Blegdamsvej 17,\\
DK-2100 Copenhagen \O, Denmark\\[0.5cm]
$^b$ Niels Bohr Institute, Blegdamsvej 17,\\
DK-2100 Copenhagen \O, Denmark
\end{center}
\vfill
\begin{abstract}
We calculate the $B_K$ parameter within the framework of the
$1/N_c$ expansion. We essentially use the technique presented
by Bardeen, Buras and G\'erard but calculate an off-shell Green
function in order to disentangle different contributions.
We study this Green function in pure Chiral Perturbation
 Theory (CHPT) first and afterwards in the $1/N_c$ expansion  in the presence
of an explicit cut-off to determine $B_K$ and the counterterms appearing
in CHPT.
The high energy part is done using the renormalization group.
For the low-energy contributions
we use both CHPT and an Extended Nambu--Jona-Lasinio
 model. This model has the right
properties to match with the high energy QCD behaviour. We then study
explicit chiral symmetry breaking effects by calculating with
 both massless and degenerate quarks together with
the real case.
A detailed analysis and comparison with the results found
within other approaches is done.
Consequences for
present lattice calculations of this parameter are then
obtained. As final result we get $0.60 < \hat B_K < 0.80$.
If $m_s=m_d=0$ we get  $0.25 < \hat B^\chi_K < 0.55$.
\end{abstract}
\vfill
February 1995
\end{titlepage}
\section{Introduction}
\rlabel{first}
\setcounter{equation}{0}

In the Standard Model (SM), strangeness changing
 processes in two units ($\Delta S=2$) can happen via
the exchange of two W-bosons as shown in Fig. \tref{Figbox}
(the so-called box diagram).
\begin{figure}
\begin{center}
\leavevmode
\epsfxsize=10cm\epsfbox{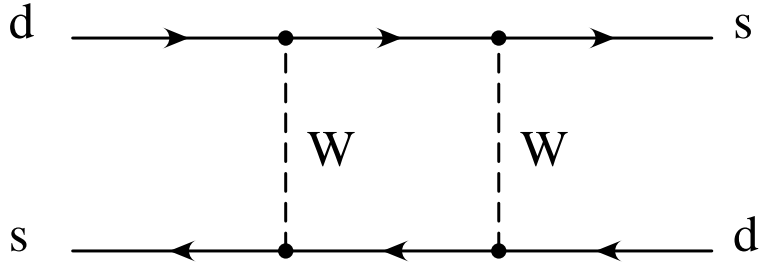}
\end{center}
\caption{The $\Delta S =2$ Box diagram}
\rlabel{Figbox}
\end{figure}
This type of transitions contributes to the
$K^0-\bar K^0$ mixing.
Experimentally this mixing is determined
by the $K_L-K_S$ oscillation length proportional to the mass difference.
The $K^0-\bar K^0$ mixing gives rise to the so-called ``indirect''
CP-violation which is usually parametrized by the CP-violating
parameter $\varepsilon$. Then the $K_L$ state consists mostly of a
CP-odd state $K_2$ with a small mixing of the CP-even state $K_1$
\be
K_L \simeq K_2 + \varepsilon K_1 ,
\ee
where $|\varepsilon| \simeq 2.3 \cdot 10^{-3}$. And the $K_S$ state
consists mostly of a CP-even state $K_1$ with a small mixing of the
 CP-odd  state $K_2$
\be
K_S \simeq K_1 + \varepsilon K_2 .
\ee
There are also contributions to this
mixing that
change strangeness in two units  through two $\Delta S=1$
transitions separated at long distances. They are important to determine the
mass difference. $\varepsilon$ is
CP-violating and is dominated by box diagram contributions.
We will concentrate on those. For an excellent recent review
on kaon CP violation see Ref. \rcite{deRAFAEL}.

At long distances, once the heaviest particles (top-quark,
$W$-boson, bottom-quark and charm-quark) have been
integrated out, the diagram in Fig. \tref{Figbox} is described by the
effective $\Delta S=2$ Hamiltonian \rcite{GAILLARD,GW83}
\be
\rlabel{hamiltonian}
{\cal H}_{\rm eff}^{\Delta S=2} (x)\,=\,
 {\cal F} \left(m_t^2,m_c^2,M_W^2,{\bf V}_{\rm CKM} \right)
\,  G_F \, \alpha_s(\mu)^{a_+} \, {\cal O}_{\Delta S=2}(x)
\ee
where ${\cal O}_{\Delta S=2} (x)$
the following $\Delta S=2$ four-quark operator
\be
{\cal O}_{\Delta S=2}(x) \equiv L^{sd}_\mu (x)
L^\mu_{sd} (x) \,
\ee
with $2 L^{sd}_\mu (x) = \bar s(x) \gamma_\mu \left(
1-\gamma_5 \right) d(x)$ and summation over colours is understood.
The strong coupling constant $\alpha_s(\mu)$ is the one with
three active light-quark flavours.
The function ${\cal F}$ is a known function depending on the
Cabibbo - Kobayashi - Maskawa (CKM) matrix elements,
 top- and charm-quark masses,
$W$ boson mass, and some QCD factors collecting the running of the
Wilson coefficients between each threshold appearing in the process
of integrating out  the heaviest particles.  For
its explicit form see Ref. \rcite{BH92,box}.
$G_F$ is the Fermi coupling constant.
The global Wilson
coefficient is dictated by the anomalous dimensions
of the operator ${\cal O}_{\Delta S=2}(x)$.
 This operator gets only multiplicatively renormalized.

The matrix element  of the operator  ${\cal O}_{\Delta
S=2}(x)$  between two on-shell kaon states is usually parametrized
in the form of
the $B_K$-parameter times the vacuum insertion approximation
(VIA) as follows
\be
\rlabel{defbk}
\langle \overline K^0  | {\cal O}_{\Delta S=2} (x)
| K^0  \rangle \equiv \frac{\dis 4}{\dis 3}
B_K (\mu) F^2_K m_K^2
\ee
where $F_K$ denotes the $K^+ \to \mu^+ \nu$ coupling ($F_K =
113$ MeV in this normalization) and $m_K$ is the $K^0$ mass.
The $\mu$-scale dependence of $B_K$ reflects the fact that
the four-quark operator ${\cal O}_{\Delta S =2}$ has an
anomalous dimension and its matrix element
depends on the scale where it is defined.
The anomalous dimension
is known and using the renormalization group
leads to the definition of the scale invariant quantity
\ba
\rlabel{bhat}
\hat B_K & =& B_K (\mu) \, \alpha_s (\mu)^{a_+}
\ea
with
\be
a_+ = \frac{\dis 3}{\dis -2 \beta^{(1)}}
\left(1-\frac{\dis 1}{\dis N_c}\right)
\ee
at one-loop. Here $\beta^{(1)}$ is the first
 coefficient of the QCD beta function.
For three active light-quark flavours and $N_c =3$ we have
$a_+ =-2/9$.
Of course, the physical matrix element
$\langle \bar K^0 | {\cal H}^{\Delta S=2}_{\rm eff} | K^0 \rangle$
is independent of the scale $\mu$. The $\mu$ dependence
of \rref{defbk} is precisely compensated in ${\cal H}^{\Delta S=2}
_{\rm eff}$ to produce a scale independent result.
 It is in this sense that $\hat{B}_K$ can be considered physical.
The anomalous dimensions and the extensions to the box diagrams needed
are known to next-to-leading logarithmic order \rcite{box}. We will
restrict ourselves to leading logarithmic order. Only this order makes sense
to the next-to-leading order in $1/N_c$ (see below) considered here.

The vacuum insertion approximation
was historically the first way this particular matrix element \rref{defbk}
was evaluated \rcite{GAILLARD}.
Here by definition we have $B_K (\mu) =1$ at any scale
and we can only obtain an order of magnitude estimate.
Next this matrix element was related to the $\Delta I=3/2$ part of
$K\to\pi\pi$ by Donoghue et al.\rcite{Donoghue1}
using SU(3) symmetry and PCAC. This leads to a value
$\hat B_K \approx 0.37$. It was then found that this relation has rather
large corrections\rcite{BSW} due to SU(3) breaking. Then three new
analytical approaches appeared, the QCD-Hadronic Duality approach
\rcite{AP1}, QCD sum rules
using three-point functions\rcite{QCD1} and
the $1/N_c$ ($N_c =$ number of colors)
expansion framework in \rcite{BBG1}. Lattice QCD
also started producing preliminary results around this time.
A review of the situation several years ago can be found in the
proceedings of the Ringberg workshop devoted to this subject\rcite{Buras1}.
All these approaches have in common that they try to get a numerical value
for the $B_K$ parameter and study its dependence on the renormalization
scale $\mu$.
All of these methods have been updated and refined. The QCD-Hadronic
Duality
update can be found in \rcite{prades1}, a QCD sum rule calculation
is in \rcite{QCD2} and the $1/N_c$ expansion method has had the vector meson
contribution calculated in a Vector Meson Dominance (VMD)
model\rcite{Gerard}. A review of recent lattice
results can be found in \rcite{lattice}. A full Chiral Perturbation
Theory (CHPT) approach to the problem is unfortunately
not possible. The data on kaon non-leptonic decays do not allow
to determine all relevant parameters at next-to-leading order
(${\cal O} (p^4)$) in
the non-leptonic chiral Lagrangian\rcite{Kambor,Ecker,Gilles}.
A calculation of these parameters within a QCD inspired model
can be found in \rcite{Bruno} where the determination of the
$B_K$ factor is done to ${\cal O} (p^4)$.

The  leading order
result for $B_K$ in the $1/N_c$ expansion is well known
\be
\rlabel{Nc}
B_K (\mu) = \hat B_K = \frac{\dis 3}{\dis 4} .
\ee
This result is model independent. However, to go further in the
$1/N_c$ expansion requires some model dependent assumptions.
Different low-energy models are then used in variants on
the $1/N_c$ method\rcite{BBG1}. An example is the calculation done
within the QCD-effective action model\rcite{AP2}.

In this paper we will use a variation on the $1/N_c$ method.
A first simplified version of this calculation has appeared in Ref.
\rcite{BP1}. There we used the Nambu--Jona-Lasinio (NJL)
 model with four-fermion
spin-1 couplings set to zero. The conclusion, there, was that although
good matching between the cut-off scale dependence from the low-energy
contribution with the perturbative QCD scale
dependence was found, that happens in the region where one
expects vector mesons to be important. We present now a complete
version with spin-1 interactions and a much more detailed
discussion of the procedure.
We use a pseudoscalar-pseudoscalar $\Delta S=2$
two-point function in the presence
of the strong interaction and the effective $\Delta S=2$
action from \rref{hamiltonian}. The method
and the reasons for this are explained in Section \tref{method}. In Section
\tref{CHPT1} we calculate this two-point function in standard Chiral
Perturbation Theory at next-to-leading order in momenta
(${\cal O}(p^4)$). This we also
use to show how the physical $B_K$ factor can be obtained from this two-point
function, and the additional information we can obtain from our method.
Here we also point out the effect of including the singlet $\eta_1$.
In the next Section, \tref{CHPTNc}, we do a first calculation of the
non-factorizable part using CHPT for the couplings. Then we give a short
overview of the extended Nambu--Jona-Lasinio model and our reasons for
using it. The main part of our work, the calculation of this two-point
function is described in Section \tref{calculation}. The checks we have on the
results are discussed next in Section \tref{checks} and finally we present
our numerical results in Section
\tref{numerics}. The conclusions from this work
are summarized in the final Section \tref{conclu}. Some examples of explicit
formulas for some of the diagrams appearing are given in an appendix.

\section{The Method and Definitions}
\setcounter{equation}{0}
\rlabel{method}

We calculate here not directly the $B_K$-factor but the
$\Delta S=2$ two-point function
\ba
\label{twopoint}
G_F\,\Pids(q^2) & \equiv &
i \int d^4 x \, e^{iq\cdot x} \, e^{i \Gamma_{\Delta S =2}}
\langle 0 | T \left( P^{ds}(0)P^{ds}(x)\right)| 0 \rangle
\nonumber \\ &
=& i^2 \int d^4 x \, e^{iq\cdot x}
\langle 0 | T \left( P^{ds}(0)P^{ds}(x) \Gamma_{\Delta S =2}
\right)| 0 \rangle
\ea
in the presence of strong interactions. We use
$P^{ds}(x) \equiv \overline d (x)i\gamma_5 s (x)$, with summation over colour
understood and
\be
\rlabel{operator}
\Gamma_{\Delta S=2} \equiv - G_F\,
\int d^4 y \, {\cal O}_{\Delta S = 2} (y) .
\ee
 The reason to calculate this two-point function rather than
directly the matrix element is that we can now perform the calculation
fully in the Euclidean region so we do not have the problem
of imaginary scalar products. This also allows us in principle to
obtain an estimate of off-shell effects in the matrix elements. This
will be important in later work to assess the uncertainty when trying
to extrapolate from $K\to\pi$ decays to $K\to 2\pi$.
This quantity is also very similar to what is used in
the lattice and QCD sum rule calculations of $B_K$.

The $\Delta S = 2$ operator in \rref{operator} can be
rewritten as
\ba
\rlabel{operator3}
\Gamma_{\Delta S=2} &=& - G_F \,
\int d^4 y L^{sd}_\mu (y) L_{sd}^\mu (y)
\nonumber \\
&=& - G_F \, \int \frac{d^4 r}{(2\pi)^4}
\int d^4 x_1 \int d^4 x_2 \,
e^{-i r \cdot(x_2 - x_1)} L^{sd}_\mu(x_1) L_{sd}^\mu(x_2) .
\ea
This allows us to consider this operator as
being produced at the $W$-boson mass scale
by the exchange of a heavy $X$ $\Delta S = 2$ boson.
So we first replace the effect of the box diagram in Fig. \tref{Figbox}
by an effective
operator of the type \rref{operator}. This then, in order to have
a physical definition of the cut-off scale, we replace by the exchange
of the X-boson. This is depicted graphically in Fig. \tref{Figreplace}.
\begin{figure}
\begin{center}
\leavevmode
\epsfxsize=12cm\epsfbox{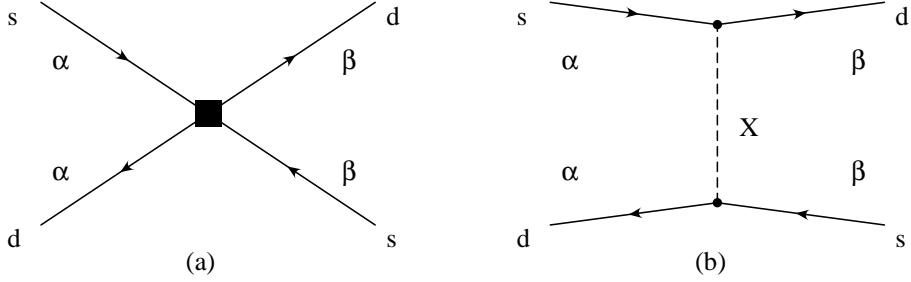}
\end{center}
\caption{(a) The operator  \protect{\rref{operator}}. (b) Its realization
via exchange of an X-boson. The colour indices that are summed over are marked
next to the quark lines.}
\rlabel{Figreplace}
\end{figure}
Notice that the identification which fermion is which, is unique in the
large $N_c$ limit.

The Feynman diagram at the quark-gluon level at leading order
in $1/N_c$ is in Fig. \tref{FigQCD}a.
\begin{figure}
\begin{center}
\leavevmode
\epsfxsize=12cm\epsfbox{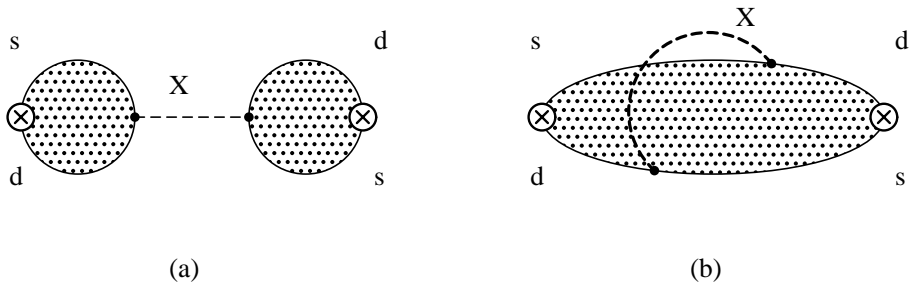}
\end{center}
\caption{The leading, (a), and the non-factorizable next-to-leading
in $1/N_c$, (b), corrections
to the $\Delta S=2$ two-point function $\Pids(q^2)$ and thus
to $B_K$.  The shaded regions are
planar QCD diagrams. The crosses
are insertions of the pseudoscalar current, $P^{ds}(x)$.}
\rlabel{FigQCD}
\end{figure}
The dotted regions are a single
quark line filled with leading in $1/N_c$ gluon exchanges, a planar diagram.
At the next-to-leading order in this expansion there are two classes of
diagrams. One is the same as in Fig. \tref{FigQCD}a but now there is
a non-planar contribution or an extra fermion loop inside each shaded region.
These we call factorizable $1/N_c$ corrections. The second class is
the diagram  shown in Fig.~\tref{FigQCD}b. Here the shaded region is
filled with gluons in a planar fashion.
It is this nontrivial class that we will compute in this paper. The first class
can be calculated completely in Chiral Perturbation Theory. They provide the
corrections needed to obtain the physical values of $F_K$, $m_K$ and
wave function renormalization.  All off-shell corrections
needed here are also purely determined by the strong interaction CHPT
coefficients.

Now we would like to give some arguments
in favour of the technique we will use. The calculation
of the hadronic matrix element in Eq. \rref{defbk} involves the mastering
of strong interactions at all energies between two very different
scales, namely, the $W$ boson mass and the kaon mass. This is, of course,
where the complexity of the calculation arises. Both
the quark-gluon momenta and the $X$-boson momentum
cover this broad range of energies. While we can use the asymptotic
freedom of QCD to perform a perturbative expansion for energies
large enough, we do not know yet how to do a QCD calculation
below energies around a few GeV. The technique we want to use
here is essentially the one used by Bardeen, Buras and G\'erard
in Ref.\rcite{BBG1,Gerard} in a slightly different notation and making
emphasis in the low-energy model to describe the strong
interactions. We want to impose as many as possible QCD
relations on this low-energy model (Weinberg Sum Rules and
similar relations).
More comments on which low-energy model we will use and why are
in Section \tref{short}.

The crucial point in this approach
is that in electroweak matrix elements
while we cannot keep track of the quark-gluon momenta
due to confinement, we can
keep track of the scale of the operator by looking at the $X$-boson
momentum. The same r\^ole
 is played by the photon in the $\pi^+ - \pi^0$ mass difference
case\rcite{BBG2,BR,BRZ}.
Looking again at the diagrams in Fig. \tref{Figreplace}
and equation \rref{operator3}, one can convince oneself
that the scale dependence imposed by the running of the quark-gluon
momenta at these energies can be identified with the dependence
on the QCD renormalization scale under some conditions.
First, for the diagrams in Fig. \tref{Figreplace} one can see that by
attaching gluon lines to the quark lines
that the dependence in a $X$-boson cut-off
can be identified  up  to corrections of order $q^2/\mu^2$
with the dependence on the gluon cut-off for these diagrams.
Here $q^2$ denotes a typical external momentum coming through the quarks.
The requirement of $q^2/\mu^2$ to be small is also the requirement that
the operator product expansion is still valid at the scale $\mu$. Otherwise
higher dimension operators are needed to be included in the
QCD running to take care of the effects of external momenta.
Of course, these would be the only corrections if the gluon propagator
had the perturbative behaviour at all scales. This we know is not the case
and is the reason why we have to go to an effective model at scales
below the chiral symmetry breaking scale $\Lambda_\chi$. The hope is then
that this effective model is sufficiently accurate up to a scale
$\mu \le \Lambda_\chi$ where both $q^2/\mu^2$ is small and the perturbative
evolution has set in. In that case a matching between the change in the
perturbative part and the change in the low-energy part with $\mu$ should
appear for some scale between $q^2$ and $\Lambda_\chi^2$.

We will work in the Euclidean domain
where all momenta squared are negative.
Then, the integral in the modulus of the momentum $r$ in
\rref{operator3} is  split into two parts,
\be \rlabel{split}
\int_0^{M_W} {\rm d} \, |r| = \int_0^\mu {\rm d} \, |r| +
\int_\mu^{M_W} {\rm d} \, |r| \ .
\ee
In principle one should then evaluate both parts separately as was done
for the $\pi^+-\pi^0$ mass difference in the above quoted references.
Notice that from the diagrams for four-quark operators at
quark-gluon level with just one-gluon line attached (i.e.
order $\alpha_s$), can only  generate logarithmically divergent terms in
a cut-off $\mu$ of the gluon momentum. One expects that the
same behaviour will appear from the low-energy
part of the integral in Eq. \rref{split} for some scale $\mu^2$
between $q^2$ and   $\Lambda_\chi^2$, as discussed before,
when the hadronic interactions are included to all
orders in momenta. Therefore, here, we will do
the upper part of the integral using the renormalization group (RG)
using the identification of the scale dependence discussed above.

An alternative way of looking at this is to
assume
that at some intermediate
scale $\mu$ we get from the low-energy calculation
\be \rlabel{gammalow}
\Gamma_{\Delta S=2} (\mu) = - G_F {\dis \int^\mu_0}
\frac{\dis {\rm d}^4 r}{\dis (2\pi)^4} \int {\rm d}^4 x_1 \, \int
{\rm d}^4 x_2 \, e^{-i r\cdot (x_2 -x _1)} \, L^{sd}_\mu
(x_1) \, L_{sd}^\mu (x_2) \, .
\ee
In QCD, the
operator $\Gamma_{\Delta S=2}(\mu)$ obeys the following inhomogeneous
renormalization group equation
\ba \rlabel{rgeq}
\mu \, \frac{\dis{\rm d} \Gamma_{\Delta S =2} (\mu)}{\dis{\rm d}\mu} &=&
- \gamma_{\Delta S=2} (\mu) \, \Gamma_{\Delta S =2} (\mu)
\nonumber \\
&=& - \left[ \gamma_{\Delta S=2}^{(1)} \frac{\alpha_s^{(1)}}{\pi}
(\mu)
+ {\cal O} \left( \left( \frac{\alpha_s}{\pi}\right)^2 \right)
\right] \Gamma_{\Delta S=2}(\mu) \, ,\nonumber \\
\gamma_{\Delta S=2}^{(1)} &=& \frac{3}{2} \left(1 - \frac{1}{N_c}
\right) \, ,
\ea
with $\gamma_{\Delta S=2}$ the gamma function of ${\cal O}_{\Delta
S=2}(x)$ \rcite{box}
and $\alpha_s^{(1)}$ the strong coupling constant to
one-loop.
This change corresponds to doing the integral in
\rref{gammalow} from $\mu$  to $\mu+{\rm d}\mu$.
Integrating then equation \rref{rgeq} between the scales $\mu$ and
$M_W$, one gets the full integral
of \rref{operator3} to be the same but integrated up to
$\mu$ and multiplied
by $C(\mu)$, with  $C(\mu)$ the corresponding Wilson coefficient.
Therefore,
\be
\rlabel{operator2}
\Gamma_{\Delta S=2} = - G_F \, C(\mu) \int_0^\mu \frac{d^4 r}{(2\pi)^4}
\int d^4 x_1 \int d^4 x_2 \, e^{-ir\cdot(x_2 - x_1)}
L^{sd}_\mu(x_1) L_{sd}^\mu (x_2)
\ee
A strict analysis in $1/N_c$ would correspond to
set $C(\mu) = 1 + d(\mu) \alpha_s^{(1)}(\mu)$
and  evaluate the second term using factorization in leading $1/N_c$.
$d(\mu)$ is what the one-gluon exchange diagrams would give with a lower
cut-off $\mu$. We will, however, use the full one-loop
Wilson coefficient $C(\mu)=\left(\alpha_s^{(1)}(\mu)/
\alpha_s^{(1)} (M_W)\right)^{a_+}$. Equation \rref{operator2},
is in fact the equivalent of the effective Hamiltonian
in Eq. \rref{hamiltonian}.
The definition of the function ${\cal F}$ includes
the factor $\alpha_s (M_W)^{-a_+}$ from the Wilson coefficient
$C(\mu)$. This permits, then, the identification
of $B_K(\mu)$.

The basic difference here with respect to the $m_{\pi^+}^2
- m_{\pi^0}^2$ case is two-fold.
First the matrix element considered
here has anomalous dimensions. It
makes the identification of the scale dependence with the
cut-off in the $X$-boson momentum  highly nontrivial. We know that
there should be an explicit logarithmic $\mu$ dependence here.
In the $\pi^+-\pi^0$ mass difference
case the matching was of order $1/\mu^2$ which
means that the intermediate momentum regime is less important.
The second difference is that in the $\pi^+-\pi^0$ case the mass difference
can be related to a vacuum matrix element\rcite{Dass}.
Here this is not possible.
So while in the other case two-point functions were sufficient we now need
to calculate four-point functions in the strong interactions.
The contribution to the $K^0- \bar K^0$ matrix element
 vanishes in the chiral limit. This, together with the fact that
the typical scale is the kaon mass makes that
 the effects of a nonzero quark mass are essential here
\footnote{In the
electromagnetic mass difference for the kaons the effects are also expected
to be large \rcite{pionmasdif}.}.
Thus  one can only  expect
matching of the $\mu$ dependence in the $B_K$ case (if any)
for scales where the effects of $nonzero$ quark masses are small, i.e.
for scales $\mu$ larger than $m_K$ .

Up to now, we did not need to specify the low-energy
model for the strong interactions. In Ref. \rcite{BBG1}, CHPT
was used, however the range of applicability of CHPT is
precisely below where one can expect a reasonable matching
with QCD (i.e., above the kaon mass). Then, in Ref. \rcite{Gerard}
vector meson interactions were included using a particular
VMD model, namely
the Hidden Gauge Symmetry model. We will calculate the
 lower part of the integral by  using an Extended
Nambu--Jona-Lasinio (ENJL) cut-off model
and also in CHPT as a test of our ENJL calculation.
Some reasons for why
do we believe this model is more suitable for this purpose
and its advantages versus other choices are in Sect. \tref{short}.
Here, we only want to point out
that ENJL  permits the control of chiral corrections. The chiral
limit is not clear in other approaches when implementing VMD, for
instance. As a matter of fact, a very important point we want to
address in this analysis, is the effect of the explicit chiral
symmetry breaking.  In addition, this model allows also for an $1/N_c$
 expansion and a chiral expansion like the one in CHPT (see
Ref. \rcite{BP2}).

\section{Chiral Perturbation Theory Calculation of $\Pids (q^2)$}
\rlabel{CHPT1}

In this Section we  study the $\Delta S=2$
two-point function $\Pids (q^2)$ in Eq. \rref{twopoint}
in the framework of  Chiral Perturbation Theory; i.e we use
CHPT to calculate the contribution of the strong interactions
at low energies.

\subsection{Lowest Order}

At lowest order in the chiral expansion ${\cal O} (p^2)$
\rcite{Weinberg}
the strong interactions between the lowest pseudoscalar mesons
including external vector, axial-vector, scalar and pseudoscalar
sources is described by the following effective Lagrangian
\be \label{L2chiral}
{\cal L}^{(2)}_{\rm eff}=\frac{F_0^2}{4} \left\{
\tr \left(D_\mu U D^\mu U^\dagger \right) +
\tr \left(\chi U^\dagger + U \chi^\dagger \right) \right\}
\ee
where  $D_\mu$ denotes the covariant derivative
\be \rlabel{covderi}
D_\mu U = \partial_\mu U - i (v_\mu + a_\mu ) U
+ i U (v_\mu - a_\mu) \, ,
\ee
and $U \equiv \exp\left( \frac{i \sqrt 2 \Phi}{F_0}\right)$
an SU(3) matrix incorporating the octet of pseudoscalar mesons
\be \rlabel{Uoctet}
\Phi(x)=\frac{\vec{\lambda}}{\sqrt 2} \vec{\phi} =
\left( \begin{array}{ccc}
\frac{\dis \pi^0}{\dis \sqrt 2} + \frac{\dis \eta_8}
{\dis \sqrt 6} &  \pi^+ & K^+ \\
\pi^- & -\frac{\dis \pi^0}{\dis \sqrt 2}+\frac{\dis \eta_8}
{\dis \sqrt 6} & K^0 \\
K^- & \overline K^0 & -\frac{\dis 2 \eta_8}{\dis \sqrt 6} \end{array}
\right) \, .
\ee
In Eq. \rref{covderi} $v_\mu(x)$ and $a_\mu(x)$ are external 3 $\times$ 3
vector and axial-vector field matrices. In Eq. \rref{L2chiral}
$\chi\equiv 2 B_0 \left({\cal M}+s(x) + i p(x)\right)$
with $s(x)$ and $p(x)$
external scalar and pseudoscalar  3 $\times$ 3 field matrices
and ${\cal M}$ the 3 $\times$ 3 flavour matrix ${\cal M}=
{\rm diag}(m_u,m_d,m_s)$ which collects the light quark masses.
The constant $B_0$ is related to the vacuum expectation
value
\be
\left. \langle 0| \bar q q | 0 \rangle \right|_{q=u,d,s} =
- F_0^2 B_0 \left(1 + {\cal O}( {\cal M})\right) \, .
\ee
In this normalization, $F_0$ is the chiral limit value corresponding
to the pion decay coupling  $F_\pi \simeq 92.5$ MeV.
In the absence of the $U(1)_A$ anomaly (large $N_c$ limit)
\rcite{Witten}, the SU(3) singlet
$\eta_1$ field becomes the ninth Goldstone boson which is
incorporated in the $\Phi(x)$ fields as
\be \rlabel{Unonet}
\Phi(x)= \frac{\vec{\lambda}}{\sqrt 2} \vec{\phi} + \frac{\eta_1}{\sqrt 3}
{\mbox {\large \bf 1}} \, .
\ee

The effective realization of the pseudoscalar current $P^{ds}(x)$ at
low-energies can be obtained from the divergence of the $ds$ component
the axial-vector quark current. Then,
\be
P^{ds}(x)= \sqrt 2 \, F_0 B_0 \bar K^0 + \cdots
\ee
where we have explicitly given the lowest order term.
The operator $\Gamma_{\Delta S=2}$ transforms as a component of a (27$_L$,
1$_R$) tensor under SU(3)$_L$ $\times$ SU(3)$_R$ chiral rotations. The
realization of this operator in terms of the relevant low-energy degrees
of freedom is determined uniquely by its symmetry structure. At leading
order in the $1/N_c$ expansion, this operator has the well-known factorizable
current $\times$ current structure
\be \rlabel{gammafac}
\Gamma_{\Delta S =2}^{1/N_c}  = - G_{27} G_F \,
{\dis \int} {\rm d}^4 y {\cal L}^\mu_{23}(y) {\cal L}_{\mu 23} (y)
\ee
with ${\cal L}_\mu = {\cal L}^{(1)}_\mu + {\cal L}^{(2)}_\mu + \cdots$
in an expansion in external momenta and quark masses. The lowest order is
${\cal L}^{(1)}_\mu = -i \frac{F_0^2}{2} U^\dagger D_\mu U$. The coupling
$G_{27}$ is a scale independent quantity modulating
the (27$_L$, 1$_R$) part under the SU(3)$_L$ $\times$ SU(3)$_R$
chiral rotations of
the Standard Model Lagrangian. In the large $N_c$ limit $G_{27}=1$.

In fact, at lowest ${\cal O}(p^2)$ in CHPT, the effective realization
of the $\Gamma_{\Delta S=2}$ operator has the factorizable structure
in Eq. \rref{gammafac} with ${\cal L}_\mu \to {\cal L}^{(1)}_\mu$.
One can now perform  the calculation of $\Pids(q^2)$
at this order. The only diagram contributing
at this order is (a) in Fig. \tref{Figloops}.
\begin{figure}
\epsfxsize=14cm
\epsfbox{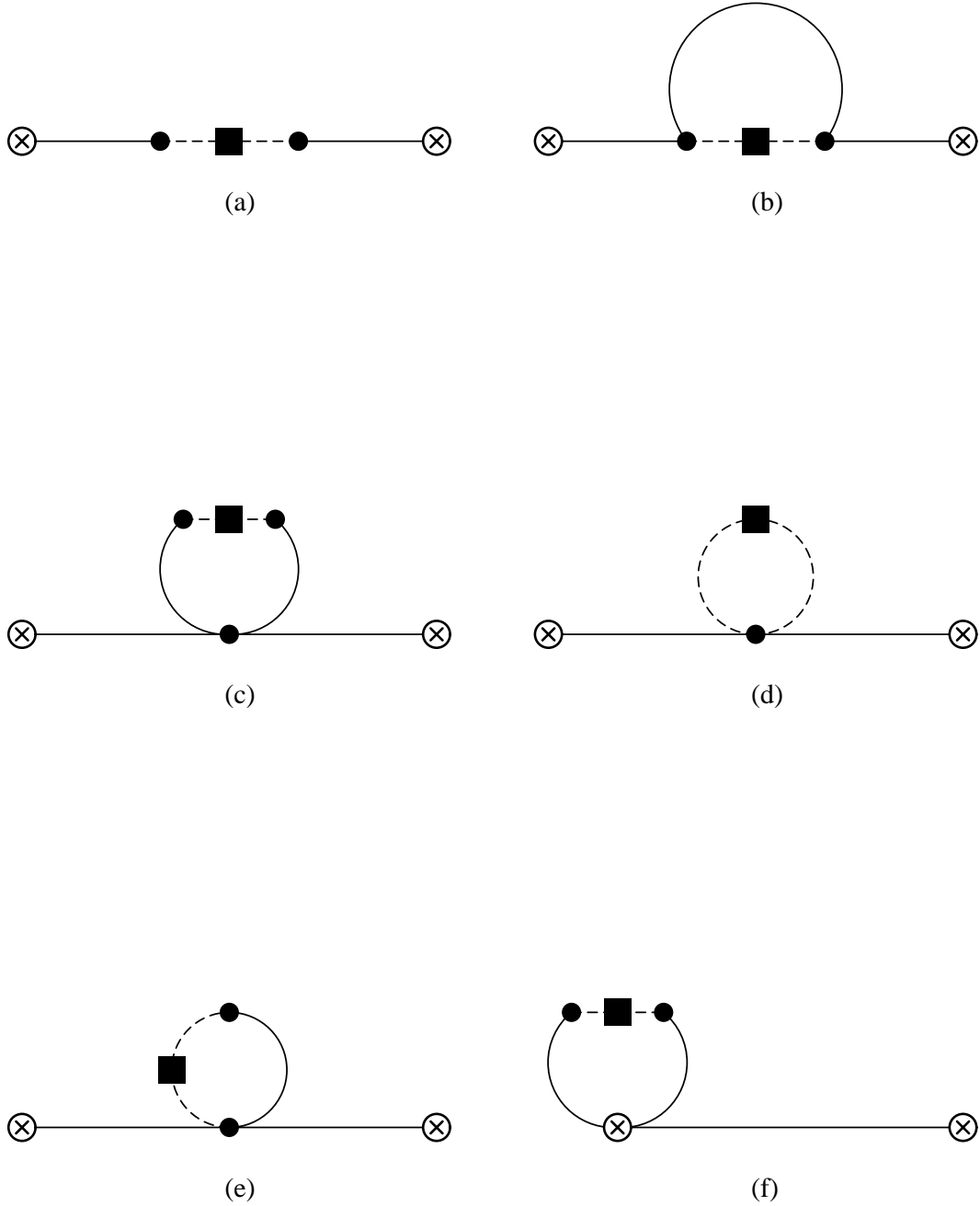}
\caption{Chiral Perturbation Theory contributions to $\Pids(q^2)$
(I) . (a) Lowest order.
(b)-(f) Higher order non-factorizable. A circled cross
is an insertion of the external pseudoscalar current, $P^{sd}(x)$,
 a dot is a strong
interaction vertex and the square
with the dashed lines to the dots represents the $\Delta S=2$ operator,
$\Gamma_{\Delta S=2}$.
The full lines are meson lines.}
\rlabel{Figloops}
\end{figure}Then, we obtain
\be \rlabel{bkchpt2}
\Pids (q^2) = - \frac{2 B_0^2 F_0^4}{\left(q^2-m_0^2\right)^2} q^2 G_{27} .
\ee
Where $m_0$ is the chiral limit value corresponding to the kaon mass.
This result is equivalent to
\be
\langle \bar K^0 | {\cal O}_{\Delta S =2}(x) | K^0 \rangle =
m_0^2 \, F_0^2 g_{27}(\mu) \qquad{\rm or}\qquad
\hat B_K^{(2)} = \frac{3}{4}G_{27} .
\ee
The coupling constant $g_{27}(\mu)$ is the overall factor modulating the
(27$_L$, 1$_R$) representation of which the operator
 ${\cal O}_{\Delta S=2}(x)$ is a component.  The relation between $G_{27}$
and $g_{27}(\mu)$ is $G_{27} \equiv g_{27}(\mu) \alpha_s(\mu)^{a_+}$.

\subsection{Next-to-Leading Order}

At ${\cal O}(p^4)$ we have to consider both tree-level
${\cal O}(p^4)$ counterterms and loops of the
${\cal O}(p^2)$ terms.
To define these loops one needs to introduce a subtraction point $\nu^2$.
This is the CHPT subtraction scale and is not related to the scale $\mu$
used in the other sections. See the more extensive discussion in Sect.
\tref{CHPTNc}.

Let us first look at the pseudo-Goldstone
boson loops. Here we will present both the results in the octet
symmetry approximation,
 i.e. using the $U$ matrix in Eq. \rref{Uoctet} with no $\eta_1$,
and in the nonet symmetry approximation (or strict large $N_c$ limit);
i.e. using the $U$ matrix in Eq. \rref{Unonet}.
First let us give the result for the octet symmetry case.
The contributions from pseudo-Goldstone boson loops
that are factorizable into two diagrams after cutting the propagator of
the fictitious $\Delta S = 2$ $X$-boson introduced in Sect.
\tref{method} can be reabsorbed in the corresponding
${\cal O}(p^4)$ expressions for $m_0$, $B_0$, and $F_0$
in Eq. \rref{bkchpt2}. They also give wave function renormalization.
These are in Figs. (a), (b) and (c) in Fig. \tref{Figloops2}
(plus the symmetric ones).

In addition, there are also non-factorizable contributions
from pseudo-Gold/-stone boson loops in Figs. (b),
(c), (e), and (f) in Fig. \tref{Figloops},
where the boxes are the $\Gamma_{\Delta S=2}$ operator
at ${\cal O}(p^2)$. The calculation of these loops gives the
following results for $\Pids(q^2)$. For the integrals we only give
the $\overline {MS}$ renormalized finite part at some scale $\nu$.
The  diagram in Fig. (b) gives
\ba
&&  i \frac{B_0^2 F_0^2}{\left(q^2 - m_K^2 \right)^2}
{\dis \int} \frac{{\rm d}^4 p}{(2 \pi)^4} \frac{1}{2}
 (q+p)^2 \left( \frac{1}{p^2-m_\pi^2} +
\frac{3}{p^2- m_{\eta_8}^2} \right)   \nonumber \\
&\doteq&  \frac{B_0^2 F_0^4}{(q^2-m_K^2)^2} \frac{1}{16 \pi^2 F_0^2} \,
\nonumber \\ &\times&
\frac{1}{2} \left[ (q^2+m_\pi^2)  m_\pi^2 \ln (m_\pi^2/\nu^2) +
3 (q^2+ m_{\eta_8}^2) m_{\eta_8}^2 \ln (m_{\eta_8}^2/\nu^2) \right] .
\ea

For the sum of diagrams (c)  and
(f) plus the symmetric one, we get
\ba
i \frac{2}{3} \frac{B_0^2 F_0^2}{\left(q^2-m_K^2\right)^2}
\, {\dis \int} \frac{{\rm d}^4 p}{(2 \pi)^4} \frac{p^2 (p^2+ 3 q^2
-m_K^2)}{\left( p^2 - m_K^2\right)^2}
\nonumber \\
\doteq  \frac{B_0^2 F_0^4}{(q^2-m_K^2)^2} \,
\frac{1}{16 \pi^2 F_0^2} \frac{2}{3}
m_K^2 \left[ (6 q^2 + m_K^2) \ln \left(m_K^2/\nu^2\right)
+3 q^2 \right].
\ea

Diagram (d) does not occur in pure CHPT.
More about this diagram is in the Sects. \tref{CHPTNc} and \tref{checks}.
Diagram (e) gives

\ba
- i \frac{8}{3} \frac{B_0^2 F_0^2}{\left(q^2 - m_K^2 \right)^2}
{\dis \int} \frac{{\rm d}^4 p}{(2 \pi)^4}
\frac{p^2}{p^2-m_K^2} \nonumber \\
\doteq  -
\frac{B_0^2 F_0^4}{(q^2-m_K^2)^2} \frac{1}{16 \pi^2 F_0^2} \,
\frac{8}{3} m_K^4 \ln (m_K^2/\nu^2)
\ea

To this result one has to add ${\cal O} (p^4)$ counterterms.
\begin{figure}
\epsfxsize=14cm
\epsfbox{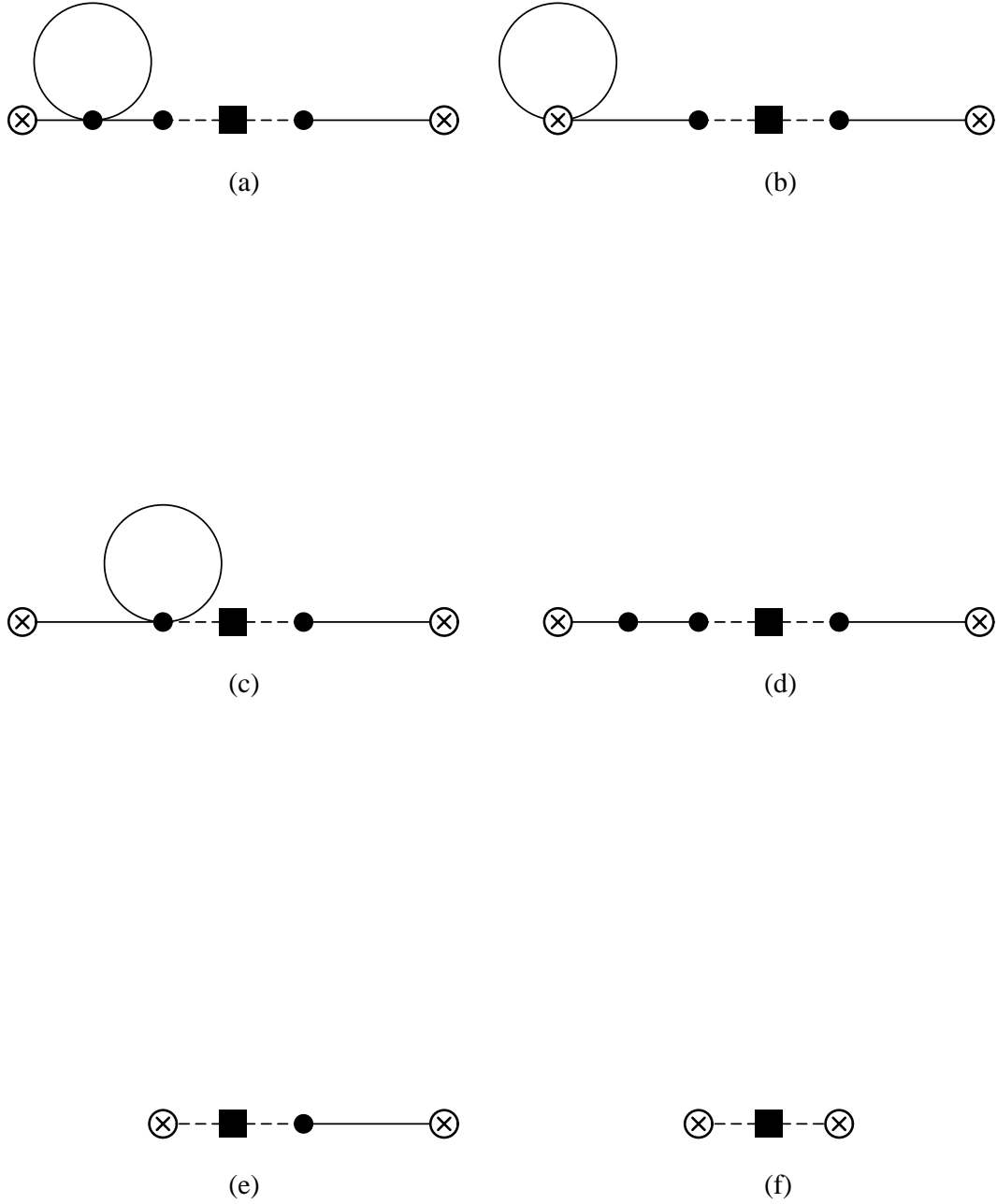}
\caption{Chiral Perturbation Theory contributions to $\Pids (q^2)$
(II) . (a)-(c) Factorizable loop corrections.
(d)-(f)  Counterterm contributions. Symbols are as in Fig.
 \protect{\tref{Figloops}}. See text for more explanation.}
\rlabel{Figloops2}
\end{figure}
The counterterms of strong origin are factorizable. The corresponding
diagrams are in Figs.
(a) in Fig. \tref{Figloops} and (d) in Fig. \tref{Figloops2}, plus the
symmetric  ones. They
 can be completely calculated from
the ${\cal O}(p^4)$  chiral Lagrangian classified by Gasser
and Leutwyler \rcite{GasserLeutwyler} in terms of
the $L_i$, $i=1, \cdots 10$ and $H_j$, $j=1,2$ couplings.
These corrections can be reabsorbed in the corresponding
${\cal O}(p^4)$ expression for $m_K$ and wave function
renormalization. In addition there are also ${\cal O}(p^4)$
$\Delta S =2$ counterterms of weak origin.  The corresponding
diagrams are in  Figs. (a) in Fig. \tref{Figloops}, (e) and (f) in
Fig. \tref{Figloops2}, plus the symmetric ones.
The general structure of these
operators was classified in Refs. \rcite{Kambor,Ecker,Gilles}.
 Then the ${\cal O}(p^4)$ $\Delta S =2$
structure of counterterms contributing to the
$\Delta S=2$ two-point function $\Pids(q^2)$ is
\ba \rlabel{counter}
\Gamma^{(4)}_{\Delta S=2} & \doteq & - G_{27} \, G_F \,
{\dis \int} {\rm d}^4 y \left[ D_1 \, F_0^4
\left(\chi^\dagger U + U^\dagger \chi \right)_{23}(y)
\left(\chi^\dagger U + U^\dagger \chi \right)_{23}(y)
\right.  \nonumber \\ &+&  D_2\, F_0^4
\left(\chi^\dagger U - U^\dagger \chi \right)_{23}(y)
\left(\chi^\dagger U - U^\dagger \chi \right)_{23}(y)
\nonumber \\ &+& i D_3 \, F_0^2 \,
{\cal L}^{(1) \mu}_{23} (y) \left( D_\mu \chi^\dagger U
- U^\dagger D_\mu \chi \right)_{23}(y)
\nonumber \\ &+&   D_4 \,  {\cal L}^{(1) \mu}_{23}(y)
\left\{\chi^\dagger U + U^\dagger \chi, {\cal L}^{(1)}_\mu
\right\}_{23}(y)  \nonumber \\
&+& i D_5 \, {\cal L}^{(1) \mu}_{23}(y)
\left[\chi^\dagger U - U^\dagger \chi, {\cal L}^{(1)}_\mu
\right]_{23}(y) \nonumber \\ &+&
 D_6 \left(\chi^\dagger  U + U^\dagger \chi \right)_{23}(y)
\left( {\cal L}^{(1)}_\mu (y) {\cal L}^{(1) \mu} (y)\right)_{23}
\nonumber \\ &+& \left. D_7 \, \tr \left( \chi^\dagger U + U^\dagger \chi
\right)(y) {\cal L}^{(1) \mu}_{23}(y) {\cal L}^{(1)}_{\mu 23}(y) \right] .
\ea
These ${\cal O}(p^4)$ $\Delta S=2$ counterterms have also a factorizable
part  which can be obtained expanding the large
$N_c$ expression in Eq. \rref{gammafac}. These
can be obtained from the corresponding ${\cal O}
(p)$ and ${\cal O}(p^3)$ terms of the left quark currents ${\cal L}_\mu$
in CHPT. Then at large $N_c$, $D_4=4 L_5/F_0^2$ and $D_7=8 L_4/F_0^2$
and all the other counterterms are zero. Here $L_5$ and $L_4$ are two
of the ${\cal O}(p^4)$ chiral Lagrangian in the strong
sector \rcite{GasserLeutwyler}. These factorizable counterterms can
once more  be absorbed in the corresponding ${\cal O}(p^4)$ expression
for $F_K$. So altogether the factorizable counterterms of both strong
and weak origin are the needed ones to cancel the (CHPT) scale dependence
generated by the factorizable  pseudo-Goldstone boson loops discussed
above, so that the factorizable part is completely calculable and
gives the result in Eq. \rref{bkchpt2} with $m_0 \to m_K$, $F_0
\to F_K$ and $(m_d+m_s) B_0 \to m_K^2$ \footnote{In fact,
this discussion can be extended to all orders in CHPT.}.

Let us discuss the non-factorizable or
 next-to-leading $1/N_c$ order contributions.
They will give the non-trivial contributions to $B_K$. Although the
calculation can be performed in terms of the couplings
$D_i$, $i=1, \cdots, 7$;  these couplings are unfortunately not known
and therefore it is not possible to  calculate
$B_K$ to ${\cal O}(p^4)$ within CHPT. Meanwhile these couplings are not
 available experimentally one can try to calculate them.
To do that one needs dynamical information for which
some QCD inspired model can be useful.  The corresponding calculation
in an effective action approach can be found in Ref. \rcite{Bruno}.
In the next sections we will use the technique and dynamical
assumptions explained in Sect. \tref{method} to obtain
in the large $N_c$ limit
these $D_i$-like couplings appearing to all orders in CHPT.

In the large $N_c$ limit we have that $D_1$, $D_2$, $D_3$,
$D_4 - 4 L_5/F_0^2$, $D_5$
$D_6$ and $D_7$ are ${\cal O}(1/N_c)$.

Diagram (a) in Fig. \tref{Figloops}
produces two kaon propagators and the result
after subtracting the part that is reabsorbed
in the factorizable or large $N_c$ contributions, is
\ba
 2 \frac{B_0^2 F_0^4}{ (q^2-m_K^2)^2} G_{27} \,
\left[ 16  D_1^r(\nu^2) \left(m_K^2 - m_\pi^2\right)^2 + 16
D_2^r(\nu^2) m_K^4
\nonumber \right. \\   \left. - q^2
\left[ 2 \left( D_7^r(\nu^2) - 8 \frac{L_4^r(\nu^2)}{F_0^2}
\right) \left( 2 m_K^2 + m_\pi^2 \right) + 4
\left( D_4^r(\nu^2) - 4  \frac{L_5^r(\nu^2)}{F_0^2}
 \right) m_K^2 \right] \right] . \nonumber \\
\ea

Diagram (e) plus the symmetric one in Fig. \tref{Figloops2}
produces one kaon propagator
and the result is
\ba
  2 \frac{B_0^2 F_0^4}{q^2-m_K^2} G_{27} \,
\, \left[ 32 D_2^r(\nu^2) m_K^2 + 4 D_3^r(\nu^2) q^2 \right] .
\ea

And diagram  (f) in Fig. \tref{Figloops2}
gives no kaon propagators and the result is
\ba
  32 B_0^2 F_0^4  G_{27} \, D_2^r (\nu^2) .
\ea
Where we have only written the $\overline{MS}$ finite part
at some scale $\nu$ in terms of $D_i^r(\nu^2)$ and $L_i^r
(\nu^2)$.

Summing up all the contributions, namely
factorizable and non-factorizable, pseudo-Goldstone boson loops
and counterterms, we get the following result for
the full  $\Pids(q^2)$ to ${\cal O}(p^4)$
\ba \rlabel{pidschpt}
\Pids (q^2) &=& - 2
\frac{ m_K^4 F_K^4}{(m_s + m_d)^2 (q^2-m_K^2)^2} G_{27} q^2
\nonumber \\ &\times& \left[ 1 +  \frac{m_K^2}
{16 \pi^2 F_0^2} \left\{ A \frac{m_K^2}{q^2} + B + C
\frac{q^2}{m_K^2} \right\} \right] .
\ea
With
\ba \rlabel{countabc}
A &=& - 16 \, [ 16 \pi^2 F_0^2 ] D_1^r(\nu^2) \,
\left( 1 - \frac{m_\pi^2}{m_K^2} \right)^2
+ \ln ( m_K^2/\nu^2) \nonumber \\
&-& \frac{1}{4} \frac{m_\pi^4}{m_K^4} \ln (m_\pi^2/\nu^2) -
\frac{3}{4} \frac{m_{\eta_8}^4}{m_K^4} \ln
(m_{\eta_8}^2/\nu^2) . \nonumber \\
B &=&  [ 16 \pi^2 F_0^2 ] \left\{ 2 \left( D_7^r
(\nu^2) - 8 \frac{L_4^r(\nu^2)}{F_0^2} \right)
\, \left(2 + \frac{m_\pi^2}{m_K^2} \right)\right.
\nonumber\\&+&
\left.  4 \left( D_4^r(\nu^2) + D_3^r(\nu^2) - 4 \frac{L_5^r(\nu^2)}{F_0^2}
 \right) \right\}
\nonumber
\\ &-& \left[ 2 \ln ( m_K^2/\nu^2) + \frac{1}{4} \frac{m_\pi^2}{m_K^2}
\ln (m_\pi^2/\nu^2) + \frac{3}{4} \frac{m_{\eta_8}^2}{m_K^2}
\ln (m_{\eta_8}^2/\nu^2)  + 1 \right]\, .
\nonumber \\
C &=& - 16 \, [16 \pi^2 F_0^2 ] \left( D_2^r(\nu^2) +
\frac{D_3^r(\nu^2)}{4} \right) .
\ea
Each of these $A$, $B$ and $C$ terms must be separately scale independent.
The divergent part of the ${\cal O}(p^4)$ couplings in Eq. \rref{counter}
was also determined in Ref. \rcite{Kambor,Ecker,Gilles}, with
\be
D_i = D_i^r(\nu^2) + \frac{\gamma_i}{16 \pi^2 F_0^2}
\nu^{d-4} \left[
\frac{1}{d-4} + \frac{1}{2} \left\{ \gamma_E - \ln (4 \pi) -1 \right\}
\right]
\ee
and $\gamma_1=-1/24$, $\gamma_2=0$, $\gamma_3=0$, $\gamma_4=3$, $\gamma_5=0$,
$\gamma_6=-3/2$, $\gamma_7=1$.
The corresponding divergent part of the ${\cal O}(p^4)$ strong chiral
Lagrangian were determined in \rcite{GasserLeutwyler} with
\be
L_i = L_i^r(\nu^2) + \frac{\Gamma_i}{16 \pi^2}
\nu^{d-4} \left[ \frac{1}{d-4}
+ \frac{1}{2} \left\{ \gamma_E - \ln (4 \pi) -1 \right\} \right] \, .
\ee
The ones we need are $\Gamma_4=1/8$ and $\Gamma_5=3/8$.
These divergences precisely cancel the ones generated by the
pseudoscalar meson loops leaving a scale independent quantity.
In the large $N_c$ counting $A$, $B$ and $C$ are ${\cal O}(1)$.
Notice that in the limit $m_s=m_d$ one gets $A =0$.
This means that the subtraction constant $\Pids(0)$ vanishes
when $m_s=m_d$.
Here we have used the Gell-Mann--Okubo octet symmetry relation
$3 m_{\eta_8}^2 = 4 m_K^2 - m_\pi^2$. The exact cancellation
of the $A$ term for equal quark masses ($m_s= m_d$)
introduces an unknown
error  in present lattice quenched calculations which are done
in this approximation.

{}From the explicit calculations one can also obtain that the
singlet
meson, $\eta_1$ does not contribute to $\Pids(q^2)$, and thus to $B_K$.
The relevant flavour
degree of freedom in the zero-charge sector is $\left[ \overline{d}d-
\overline{s}s \right]$ which is octet.
So only SU(3) breaking effects
can be important. In order to estimate these we have also
performed the calculation in the nonet symmetry approximation;
i.e. using the $U$ matrix in Eq. \rref{Unonet}. This is the strict
large $N_c$ limit and is the one analogous to the calculation  we shall
do afterwards using the ENJL as low-energy hadronic model. Also,
as in the ENJL model calculation, we have
taken  the $\bar q q$ states as basis for the states running in the loops
and not the mesonic basis of the lowest
pseudoscalar mesons, namely  $\pi$, $K$ and $\eta$ mesons.
This will change some of the
scale dependence of the logarithms since the dynamical fields are
not the same. In fact, from the calculation using nonet symmetry and
$\bar q q$ states we get the same expression as in Eq. \rref{pidschpt}
but changing $m_{\eta_8}^{2(4)} \ln ( m_{\eta_8}^2 / \nu^2)$
by $(2/3) \, ( 2 \, m_K^2 - m_\pi^2 )^{1(2)}
 \ln ( (2 \, m_K^2 - m_\pi^2)/\nu^2 ) $
and $m_\pi^{2(4)} \ln (m_\pi^2/\nu^2) $
by $2 m_\pi^{2(4)} \ln (m_\pi^2/\nu^2)$.
The exponent in brackets is for the change in $A$, the other one is for the
change in $B$. We have used here that the $\overline{d}d$ meson has the same
mass as the pion while the $\overline{s}s$ one has a mass
$2m_K^2 - m_\pi^2$.
This produces that in this basis of fields the divergences are
 $\gamma_1=-1/8$ and the remaining ones the same as given above.

The different chiral logarithms in the
octet case versus the nonet case
for the $A$ and $B$ terms produces a numerical difference, which
for a scale $\nu=m_\rho \simeq 0.77$ GeV, is
\be
A_{\rm octet} = A_{\rm nonet} + 0.36
\ee
and
\be
B_{\rm octet} = B_{\rm nonet} + 0.32 \ .
\ee

Let us now go to the $B_K$ parameter  itself. After reducing the
$\Delta S=2$ two-point function above we get to ${\cal O}(p^4)$
\ba
\hat B_K^{(4)} &=& \frac{3}{4} G_{27} \,
\left[ 1 + \frac{m_K^2}{16 \pi^2 F_0^2} \left\{ A + B + C \right\}
\right] .
\ea
The difference obtained above for the $A$ and $B$ terms from
the chiral logarithms in the octet case versus the nonet case enlarge
$\hat B_K^{(4)}$ by 0.09 in the octet case, i.e. no $\eta_1$.

\subsection{Large $N_c$ expansion}

In this subsection we would like to perform also an $1/N_c$
discussion in the framework of CHPT without additional
dynamical  assumptions.  As said before in the
large $N_c$ limit the $\Gamma_{\Delta S=2}$ operator has
the factorizable structure in Eq. \rref{gammafac}.
We also know that in the large $N_c$ limit $G_{27}=1$. So
to all orders in CHPT and leading $1/N_c$
\be \rlabel{bknc}
\hat B_K = \frac{3}{4} \, .
\ee

The next-to-leading order $1/N_c$ corrections to this result
are more involved.
At lowest order CHPT ${\cal O}(p^2)$ one still has only the factorizable
structure at next-to-leading ${\cal O} (1/N_c)$
when substituting ${\cal L}_\mu$ by ${\cal L}_{\mu}^{(1)}$ in Eq.
\rref{gammafac}. This happens to all orders in $1/N_c$
and lowest ${\cal O}(p^2)$.
Now $G_{27}$ contains $1/N_c$ corrections. Then to
${\cal O}(p^2)$ and to all orders in the $1/N_c$ expansion
\be
\hat B_K^{(2)} = \frac{3}{4} G_{27} .
\ee
Donoghue et al. \rcite{Donoghue1} used SU(3) chiral symmetry
to estimate $G_{27}$ from the measured $K^+ \to \pi^+ \pi^0$
$\Delta I =3/2$ decay rate which is modulated by the same $G_{27}$.
They estimated it to be
\be \rlabel{G27B}
G_{27} \simeq 0.88 \frac{F_0^2}{F_K^2} \simeq 0.49
\ee
and
\be \rlabel{bkp2}
B_K^{(2)} \simeq 0.37 \, .
\ee
These two results above in \rref{bknc} and \rref{bkp2} are quite
well established. Now, one can go to ${\cal O}(p^4)$ and next-to-leading
$1/N_c$. Again one has the factorizable contributions that are reabsorbed
in the values of $F_K$, $m_K$ and in wave function renormalization.
But at this ${\cal O}(p^4)$,
there are also non-factorizable $1/N_c$ corrections.
The result at next-to-leading ${\cal O}(1/N_c)$ can be obtained
from the previous section,
\ba
\hat B_K^{(4)} &=& \frac{3}{4} \,
\left[ G_{27} + \frac{m_K^2}{16 \pi^2 F_0^2} \left\{ A + B + C \right\}
\right] .
\ea
Where now $G_{27}$ contains $1/N_c$ corrections and the $A$, $B$ and $C$
terms have the same expression but calculated
to ${\cal O}(1)$ in the large $N_c$.

{}From the $A$, $B$ and $C$ terms one can see that there are
five structures of counterterms to be determined. Namely, $G_{27}$,
$D_1^r$, $D_7^r - 8 L_4^r/F_0^2$, $D_4^r + D_3^r - 4 L_5^r/F_0^2$ and
$D_2^r + D_3^r/4$ \footnote{Notice that the combinations $D_4+D_3$ and
$D_2 + D_3/4$ cannot be disentangled.
The operator multiplied by $D_3$ can be rewritten in terms of the
others using field transformations.}
To do that we need dynamical information on the strong interactions.
In the next section we will see how they can be determined in the
approach explained in Sect. \tref{method} using CHPT to next-to-leading
order. Then, in Sect. \tref{numerics} we will use the full ENJL to
obtain the low-energy contribution to the $\Gamma_{\Delta S=2}(\mu)$
function. There,  the dynamical assumptions are both in the use
of the ENJL model and in the identification of the cut-off scale
$\mu$ with the perturbative renormalization scale. We are in this case
calculating the $\Delta S=2$ two-point function
at next-to-leading order in the $1/N_c$ expansion and to all orders
in CHPT. So, we are in fact calculating all the $D_i$-like counterterms
that appear to all orders in CHPT at leading ${\cal O}(1/N_c)$.

Let us now sketch how this can be done to ${\cal O}(p^4)$.
In the chiral limit  there are two counterterms
\be \rlabel{chiralcount}
G_{27} \hspace*{2cm} {\rm and}
\hspace*{2cm} C = D_2^r + \frac{D_3^r}{4} \, .
\ee
In the chiral limit and on-shell there is only $G_{27}$.
Then we can obtain $G_{27}$ from the chiral limit for $q^2=0$.
The slope in this case will give the $C$ term.
Outside the chiral limit there are three more structures to be determined.
These can be determined due to their different $q^2$ behaviour.
The $A$ term produces a pole in the $B_K(\mu^2,q^2)$ form factor
at $q^2=0$ for $m_d \neq m_s$. From its  residue on can solve for
the $D_1$ counterterm. The $B$ term has two possible structures
involving $D_7$ and $D_4$ respectively, that can be disentangled
from the different $m_\pi^2/m_K^2$ dependence multiplying them.

In Sect. \tref{numerics} we will see how an analogous discussion can
be done to all orders in CHPT. There we will first
determine the $G_{27}$ coupling
by going to the chiral limit and making a numerical fit to
to the ratio between the lowest order to the next-to-leading
$1/N_c$ order to a  polynomial in powers of $q^2$
starting by $q^0$.  Then $G_{27}$ is the term at $q^2=0$.
Once we have $G_{27}$,
we do both, for the case of degenerate
 quarks $m_s=m_d$ and the real case $m_s \neq m_d$,
another numerical fit to the ratio between the lowest
order to the next-to-leading $1/N_c$ order to a  polynomial
in powers of $q^2$ starting in $q^{-2}$.
Here, we are determining the $A$, $B$, $C$, $\ldots$ terms; i.e.
the $D_i$-like counterterms.

We have one more general result in CHPT. To next-to-leading order
in $1/N_c$ but to all orders in CHPT the diagrams
 that can contribute are still
those depicted in Figs. \tref{Figloops} and \tref{Figloops2}. The vertices now
are the ones appearing in the CHPT Lagrangian at all orders. There is still
no non-analytic dependence on $q^2$ possible even with all possible vertices.
The result from the previous section that at order $p^4$ the dependence on
$q^2$ in $\Pids$ is analytic, is thus true to all orders up to
next-to-leading order in $1/N_c$.

\section{Chiral Perturbation Theory Calculation
in the $1/N_c$ Expansion}
\setcounter{equation}{0}
\rlabel{CHPTNc}

In this section we use CHPT to estimate the low-energy contribution
to $\Gamma_{\Delta S=2}(\mu)$ in Eq. \rref{gammalow}.
We will only discuss the nonet symmetry case and using
the states $\bar q q$ as dynamical basis for the states running in the
loops as explained in the previous section.
As said in Section \tref{method} at some intermediate energy
region we want to identify the dependence on a cut-off $\mu$ in
the $X$-boson momentum with the QCD renormalization scale dependence.
Under the conditions explained there, one expects it to be plausible for
some value below the  spontaneous symmetry breaking scale.

This scale $\mu$ is thus not related
to the scale $\nu$ of the previous section,
even though it appears in a similar fashion in the logarithms. If we
wanted to extend the analysis presented here going beyond the
lowest order in CHPT coupling of currents to the mesons
a similar scale $\nu$ would appear. This would then cancel the
dependence on $\nu$ of the $L_i$ in the higher order CHPT Lagrangian.
The answer
then would be $\nu$ independent. The scale $\mu$ is the upper limit
of the integral in Eq. \rref{gammalow} and would still be present.

We have chosen to route the momentum in the loop integrals $r$
as $p_X = r +q$ where $p_X$ is the $X$-boson momentum and $q$ is the
external momentum.  Any other routing will induce similar uncertainties
of ${\cal O}(q^2/\mu^2)$.
Then, we do these integrals in the Euclidean
space cutting-off the loop momentum $r_E$  for $|r_E|^2 > \mu^2$
(where the subscript $E$ stands for Euclidean).

We want to emphasize here that this is not a pure CHPT
calculation as in Section \tref{CHPT1}.
 It contains some dynamical assumptions
like that we can reproduce the QCD renormalization scale
dependence  in the $1/N_c$ expansion
with a cut-off in the $X$-boson momentum, i.e. that lowest order CHPT is
good enough up to a scale $\mu$ where we can compare with the perturbative
part of the calculation.
We will make then both  an expansion in $q^2/ \Lambda_\chi^2$,
$\mu^2/ \Lambda_\chi^2$ and quark masses over the same scale
$\Lambda_\chi^2$ in the context of an $1/N_c$ expansion.
$\Lambda_\chi \simeq 1.2$~GeV is
the scale of spontaneous symmetry breaking. In the previous section
the requirement was $q^2/\Lambda_\chi^2$ and $m_K^2/\Lambda_\chi^2$ small.

In addition to being an estimate of the non-leptonic parameters in the sense
of Ref. \rcite{BBG1},
CHPT provides a model independent result that will help us to
check our ENJL calculation for $q^2/ \Lambda_\chi^2$ and
$\mu^2/ \Lambda_\chi^2$ small enough.

In this notation of CHPT the QCD quark current $P^{ds}(x)$ couples to the
external source $[p(x)]_{23}$ and the left current $L_\mu^{sd}(x)$ to
$[v_\mu(x) - a_\mu(x)]_{32}$.

At lowest order in the chiral expansion only the factorizable
diagram in Fig. \tref{Figloops}
contributes. This contribution is ${\cal O} (N_c^2)$
in the  $1/N_c$ expansion. The result is
\be \rlabel{LOCHPT}
-\frac{2 B_0^2 F_0^4}{\left(q^2-m_0^2\right)^2} \, q^2\, .
\ee
This result when properly reduced is equivalent to
\be
\langle \overline K^0 | {\cal O}_{\Delta S=2}(x) | K^0 \rangle
  = m_0^2 F_0^2 .
\ee

At the next order
we have both factorizable ${\cal O} (N_c^2)$ and non-factorizable
${\cal O}(N_c)$ contributions. The ones that are factorizable are
given in terms of  the couplings of the ${\cal O}(p^4)$
chiral Lagrangian which was classified by Gasser and Leutwyler
and loops of the ${\cal O}(p^2)$ chiral Lagrangian.
These are completely calculable and we do not need any dynamical
assumption, they only involve momenta of order $q^2$
and not of order $\mu^2$.
These corrections are precisely the ones that redefine the
$B_0^2 F_0^4$ couplings to its
${\cal O}(p^4)$ expressions and are shown
in Fig. (a) in Fig. \tref{Figloops}  and Figs. (a) to (d)
plus the symmetric ones in  Fig. \tref{Figloops2}.
Those that are non-factorizable are shown in Figs. (b) to (f)
plus the symmetric ones in Fig. \tref{Figloops}.

Diagram (b) in Fig. \tref{Figloops}
gives the following contribution
\ba \rlabel{firstdiag}
-\frac{B_0^2 F_0^2}{(q^2-m_K^2)^2} {\dis \int^\mu}
\frac{{\rm d}^4 r_E}{(2\pi)^4} \, (r_E-q_E)^2 \left[\frac{\dis 1}
{\dis r_E^2 + m_\pi^2} + \frac{\dis 1}{\dis r_E^2 + 2 m_K^2 - m_\pi^2}
\right] \nonumber \\
= - \frac{B_0^2 F_0^4}{(q^2-m_K^2)^2} \, \frac{1}{16 \pi^2 F_0^2}
\, \left[ \mu^4 - (q^2 + m_\pi^2) \left( \mu^2 - m_\pi^2
\ln \left( \frac{\mu^2 - m_\pi^2}{m_\pi^2} \right) \right)
\right. \nonumber \\
\left. - (q^2 + 2 m_K^2 - m_\pi^2 ) \left( \mu^2 - \left( 2 m_K^2
- m_\pi^2 \right) \ln \left( \frac{\mu^2 + 2 m_\pi^2 - m_\pi^2 }{2 m_K^2
- m_\pi^2} \right) \right) \right] \, .
\ea
Keep in mind that the mass of the $\overline{s}s$ state is $2m_K^2-m_\pi^2$.

The sum of diagrams (c) and (f) plus the symmetric one
in Fig. \tref{Figloops} gives
\ba \rlabel{seconddiag}
&& -\frac{2}{3}
\, \frac{B_0^2 F_0^2}{(q^2-m_K^2)^2} {\dis \int^\mu}
\frac{{\rm d}^4 r_E}{(2\pi)^4} \, \frac{(r_E+q_E)^2 \left[
3 q_E^2 + (r_E+q_E)^2 + m_K^2\right]}{\left( (r_E+q_E)^2
+ m_K^2 \right)^2} \nonumber \\
&=& - \frac{B_0^2 F_0^4}{(q^2-m_K^2)^2} \, \frac{1}{16 \pi^2 F_0^2}
\nonumber \\
&\times& \frac{2}{3}
\, \left[ \frac{\mu^4}{2} - \left( 3 q^2 + m_K^2 \right) F_1(\mu^2,q^2,m_K^2)
  +3q^2 m_K^2 F_2(\mu^2,q^2,m_K^2)\right]
\nonumber\\
&=_{\displaystyle m_K^2\to 0}&
 - \frac{B_0^2 F_0^4}{(q^2-m_K^2)^2} \, \frac{1}{16 \pi^2 F_0^2}
\frac{1}{3}\left[ \mu^4-6q^2\mu^2-3q^4\right]
\nonumber\\
&=_{\displaystyle q^2\to 0}&
 - \frac{B_0^2 F_0^4}{(q^2-m_K^2)^2} \, \frac{1}{16 \pi^2 F_0^2}
\frac{1}{3}\left[ \mu^4 -  2 m_K^2 \left( \mu^2 -
m_K^2 {\rm ln}\left(\frac{\mu^2+m_K^2}{m_K^2}\right) \right)
\right]\ .
\nonumber\\
\ea
The functions $F_1$ and $F_2$ are \be
F_{i}(\mu^2,q^2,m_K^2) = \int_0^{\mu^2}{\rm d}s\frac{1}{2q^2}
\lambda^{2-i}\,\left[1-\frac{s+m_K^2-q^2}{\lambda}\right]\ ,
\ee
using $\lambda = \sqrt{(s+m_K^2-q^2)^2+4s q^2}$.
These integrals can be performed analytically but the result
 is very cumbersome.
The final two lines in Eq. \rref{seconddiag}
give the expressions for $m_K^2=0$ and $q^2=0$,
respectively.

Diagram (d) in Fig. \tref{Figloops}
gives no contribution at this order. This diagram deserves
more attention related with the Fierzed terms \rcite{BP1}.
It provides in fact another non-trivial check. In the soft pion
limit, one can  relate the VV (AA) part of this diagram with a sum rule
to some moment of a VV-AA two-point function
\rcite{DG93}.  These issues are discussed in Section \tref{checks}.

Diagram (e) in Fig. \tref{Figloops} gives
\ba \rlabel{thirddiag}
\frac{8}{3}
\frac{B_0^2 F_0^2}{(q^2-m_K^2)^2} {\dis \int^\mu}
\frac{{\rm d}^4 r_E}{(2\pi)^4} \, \frac{(r_E+q_E)^2}
{(r_E+q_E)^2 + m_K^2 }
\nonumber \\
= \frac{B_0^2 F_0^4}{(q^2 - m_K^2)^2} \frac{1}{16 \pi^2 F_0^2}
\frac{8}{3} \left[ \frac{\mu^4}{2} - m_K^2 F_1(\mu^2,q^2,m_K^2)\right]
\nonumber \\
\doteq \frac{B_0^2 F_0^4}{(q^2 - m_K^2)^2} \frac{1}{16 \pi^2 F_0^2}
\frac{8}{3} \left[ \frac{\mu^4}{2} -
m_K^2 \left( \mu^2 - m_K^2 \ln \left( \frac{\mu^2 + m_K^2}{m_K^2}
\right) \right) \right] \, .
\ea
The result in the last line above is exact
both for $q^2 =0$ and for $m_K^2=0$.

The quartic dependence in the cut-off
cancels between the different diagrams as required
by chiral symmetry. The $1/N_c$ expansion and the
dynamical assumptions mentioned above have allowed us
to make a full calculation of the two-point function
$\Pids(q^2)$ both at next-to-leading ${\cal O}(N_c)$
in the $1/N_c$ expansion  and ${\cal O}(p^4)$ in the chiral expansion.
The factorizable loops have a $\nu$ dependence that cancels as explained
in the beginning of this section.
The integrals in \rref{firstdiag}, \rref{seconddiag} and
\rref{thirddiag} produce both analytical and non-analytical
$\mu$-scale dependence. The cut-off procedure we followed
has produced  logarithmic dependence in $\mu$ for
$B_K$ proportional
to meson masses  (see below). This dependence should cancel when the
chiral expansion for $\Pids(q^2)$ is considered to all orders
giving no contribution to the running of $B_K(\mu)$ for large $\mu$.
The perturbative running is after all independent of the quark masses.
Lowest order CHPT as used here is expected to loose validity before
we reach such scales.

Here, we are actually calculating the combination of
counterterms and chiral logs in
Eqs. \rref{chiralcount} and \rref{countabc}. So in order to extract
the values of the coupling constants we should compare the full expression
for $\Pids(q^2)$ in  \rref{pidschpt}, with the one calculated here.

In the chiral case $m_s=m_d=0$
 the calculation of the next-to leading
order contribution to $\Pids(q^2)$
can be written as (assuming $\mu^2 > q^2$)
\be \rlabel{NLOCHPT1}
\frac{ B_0^2 F_0^4}{\left(q^2-m_K^2\right)^2}
\frac{1}{16 \pi^2 F_0^2} \left[ 4 \mu^2 q^2 +  q^4 \right] \, .
\ee
This result added to the one in \rref{LOCHPT}, then
properly reduced and compared with the
$B_K$ definition in Eq. \rref{defbk} gives
\be
B_K(\mu^2,q^2) = \frac{3}{4} \left(1-\frac{1}{16 \pi^2 F_0^2}
\left[ 2 \mu^2 + \frac{q^2}{2}\right] \right) .
\ee
As we already knew from the CHPT calculation in
Sect. \tref{CHPT1} there is no pole for this case.
 Doing now the general $m_s
\neq m_d$ case we can get the $G_{27}$ coupling and the
$A$, $B$ and $C$ terms. Then to next-to-leading order
in $1/N_c$ and to order $p^4$, i.e. we have neglected
the higher order CHPT terms in the last line
in Eq. \rref{seconddiag},
\ba
\rlabel{chptresult}
G_{27} &=&   \alpha_s(\mu)^{a_+}
- \frac{\mu^2 }{8 \pi^2 F_0^2}  \nonumber \\
A &=&  -
\ln \left( \frac{\mu^2 + m_K^2}{m_K^2} \right)
+ \frac{1}{2}  \frac{m_\pi^4}{m_K^4} \ln \left(
\frac{\mu^2+m_\pi^2}{m_\pi^2}\right)  \nonumber \\
&+&  \frac{1}{2} \left(2 - \frac{m_\pi^2}{m_K^2} \right)^2
\ln \left( \frac{\mu^2+2m_K^2-m_\pi^2}{2m_K^2-m_\pi^2} \right)
 \nonumber \\
B &=&
2 \ln \left( \frac{\mu^2+ m_K^2}{m_K^2} \right)
+\frac{1}{2} \frac{m_\pi^2}{m_K^2}
\ln \left( \frac{\mu^2 + m_\pi^2}{m_\pi^2} \right)
 \nonumber \\ &+&
 \frac{1}{2} \left( 2 -  \frac{m_\pi^2}{m_K^2} \right)\,
\ln \left( \frac{\mu^2 +
2 m_K^2 - m_\pi^2}{2 m_K^2 - m_\pi^2 } \right)
+ {\cal O} \left( \frac{m_K^2}{\mu^2} \right) . \nonumber \\
C &=& - \frac{1}{2} +
{\cal O} \left(\frac{m_K^2}{\mu^2} \right)  \, .
\ea
Clearly the matching between  the QCD perturbative
scale dependence and the cut-off $\mu$
here in $G_{27}$, $A$, $B$ and $C$
is not good enough. Remember they have to be separately scale independent.
To improve this situation we need to go to higher order in CHPT.
This we will do in the next sections. Using an ENJL model as a good
low-energy hadronic model we will calculate to all orders in CHPT
in the framework of the $1/N_c$ expansion. This ${\cal O}(p^4)$
will serve as a check of our ENJL calculation. In fact, the
comparison of this result ($G_{27}$, $A$, $B$, and $C$)
with the ENJL calculation
for $q^2/\Lambda_\chi^2$ and $\mu^2/\Lambda_\chi^2$ small
enough is good.

Taking the point of view of Ref. \rcite{BBG1} we can compare Eqs.
\rref{chptresult} and \rref{countabc} to obtain
\ba
D_1^r(\nu^2)=D_7^r(\nu^2)-\frac{8L_4^r(\nu^2)}{F_0^2} &=& 0\ ,\nonumber\\
\left[ 16 \pi^2 F_0^2 \right] \left( D_4^r(\nu^2) + D_3^r(\nu^2)-
\frac{4L_5^r(\nu^2)}{F_0^2} \right) &=& \frac{1}{4}\ ,
\nonumber\\
\left[ 16 \pi^2 F_0^2 \right]  \left( D_2^r(\nu^2)+\frac{D_3^r(\nu^2)}{4}
\right) &=& \frac{1}{32}\ .
\ea
Notice that this assumes that for $\mu^2 >> m_K^2$, this ${\cal O}(p^4)$
calculation is enough to reproduce exactly the QCD perturbative scale
dependence and thus one can choose $\mu=\nu$. This assumption, if good
anywhere, would be good for scales $\mu$ around or below the spontaneous
symmetry breaking scale ($ \simeq 1.2$ GeV).

\section{Short Description of
the ENJL Model and its Connection with QCD}
\rlabel{short}
\setcounter{equation}{0}

For recent comprehensive reviews on the NJL model \rcite{NJL}
and the ENJL model \rcite{ENJL}, see Refs. \rcite{Reviews,Physrep}.
Here, we only will give a brief introduction and reasons why we choose
this model.
The QCD Lagrangian is given by
\ba
\rlabel{QCD}
{\cal L}_{\rm QCD} &=& {\cal L}^0_{\rm QCD} -\frac{1}{4}G_{\mu\nu}
G^{\mu\nu} \, , \nonumber\\
{\cal L}^0_{\rm QCD} &=& \overline{q} \left\{i\gamma^\mu
\left(\partial_\mu -i v_\mu -i a_\mu \gamma_5 - i G_\mu \right) -
\left({\cal M} + s - i p \gamma_5 \right) \right\} q \, .
\ea
Here summation over colour degrees of freedom
is understood and
we have used the following short-hand notations:
$\overline{q}\equiv\left( \overline{u},\overline{d},
\overline{s}\right)$; $G_\mu$ is the gluon field in the
fundamental SU(N$_c$) (N$_c$=number
of colours) representation;
$G_{\mu\nu}$ is the gluon field strength tensor in
the adjoint SU(N$_c$) representation; $v_\mu$, $a_\mu$, $s$ and
$p$ are external vector, axial-vector, scalar and pseudoscalar
field matrix sources; ${\cal M}$ is the quark-mass matrix.

{}From lattice QCD numerical simulations, all indications are that in
the purely gluonic sector there is a mass-gap. Therefore there seems to be
a kind of cut-off mass in the gluon propagator (see the discussion in
Ref. \rcite{LatticeNJL}).
Alternatively one can think of integrating out the
high-frequency (higher than $\Lambda_\chi$, a cut-off of the order
of the spontaneous symmetry breaking scale) gluon
and quark degrees of freedom and then expand the resulting effective
action in terms of local fields.
We then stop this expansion after the dimension
six terms. This leads to the following effective action
at leading order in the $1/N_c$ expansion
\ba
\rlabel{ENJL1}
{\cal L}_{\rm QCD} &\rightarrow&  {\cal L}_{\rm ENJL} =
{\cal L}_{\rm QCD}^{\Lambda_\chi}
+ {\cal L}_{\rm ENJL}^{\rm S,P} + {\cal L}_{\rm ENJL}^{\rm V,A} +
{\cal O}\left(1/\Lambda_\chi^4\right), \\
\rlabel{ENJL2}
{\rm with}\hspace*{1.5cm}
{\cal L}_{\rm ENJL}^{\rm S,P}&=&
\frac{\dis 8\pi^2 G_S \left(\Lambda_\chi \right)}{\dis
N_c \Lambda_\chi^2} \, {\dis \sum_{i,j}} \left(\overline{q}^i_R
q^j_L\right) \left(\overline{q}^j_L q^i_R\right) \\
\rlabel{ENJL3}
{\rm and}\hspace*{1.5cm}
{\cal L}_{\rm ENJL}^{\rm V,P}&=&
-\frac{\dis 8\pi^2 G_V\left(\Lambda_\chi\right)}{\dis
N_c \Lambda_\chi^2}\, {\dis \sum_{i,j}} \left[
\left(\overline{q}^i_L \gamma^\mu q^j_L\right)
\left(\overline{q}^j_L \gamma_\mu q^i_L\right) + \left( L \rightarrow
R \right) \right] \,. \nonumber \\
\ea
Here $i,j$ are flavour indices and $\Psi_{R,L} \equiv
(1/2) \left(1 \pm \gamma_5\right) \Psi$.
The couplings $G_S$ and $G_V$ are
dimensionless and ${\cal O}(1)$ in the $1/N_c$ expansion and summation
over colours between brackets is understood.
The Lagrangian ${\cal L}^{\Lambda_\chi}_{\rm QCD}$ includes only
low-frequency modes of quark and gluon fields.
The remaining
gluon fields can be assumed to be fully
absorbed in the coefficients of the local quark field operators
or alternatively also described
by vacuum expectation values of gluonic operators.
So at this level we have two different pictures of this model. One is where
we have integrated out all the gluonic degrees of freedom and then
expanded the resulting effective action
in a set of {\bf local} operators keeping only the first nontrivial
terms in the expansion.
In addition to this we can make additional assumptions.
If we simply assume that these operators are produced by the short-range part
of the gluon propagator we obtain $G_S = 4 G_V = N_c\alpha_s/\pi$.
The
two extra terms in \rref{ENJL2} and \rref{ENJL3}
have however different anomalous dimensions
so at the strong interaction regime where these should be generated there is
no reason to believe this relation to be valid. In fact the best fit
to the low energy effective chiral Lagrangians of ${\cal O}(p^2)$
and ${\cal O}(p^4)$ \rcite{BBR}, is for $G_S \simeq 1.216\approx G_V
\simeq 1.263$ (Fit 1 in Ref. \rcite{BBR}).
The third parameter appearing in this picture
is also obtained from the same fit, with  $\Lambda_\chi=1.16$ GeV.
The light quark masses in ${\cal M}$ are fixed then to
obtain the physical pion and kaon masses in the poles of the
pseudoscalar two-point functions \rcite{BP2}.
Then we get $\overline m \equiv (m_u+m_d)/2 =3.2$ MeV and $m_s=83$ MeV.

The other picture is the one where we only integrate out the short distance
part of the gluons and quarks. We then again expand the resulting effective
action in terms of low-energy gluons and quarks in terms of local
operators. Here we make the additional assumption that gluons only
exists as a perturbation on the quarks. The quarks feel only the interaction
with background gluons. This is worked out by only keeping the vacuum
expectation values (VEVs) of gluonic operators and not including propagating
gluonic interchanges. Most fits are in fact better with this
gluonic VEV set to zero and  when this is not so
the results are quantitatively very close to the results in that case.
Accordingly, we will take this gluonic VEV equal to zero in this work.

This model has the same symmetry structure as the QCD action
at leading order in $1/N_c$ \rcite{tHooft}. Notice that the $U(1)_A$
problem is absent at this order \rcite{Witten}.
(For explicit symmetry properties under SU(3)$_L$ $\times$
SU(3)$_R$ of the fields in this model
see reference \rcite{BBR}.) The QCD chiral flavour
anomaly can also be consistently reproduced in these
kind of models when spin-1 four-fermion  interactions are included
\rcite{BP3}.
In the chiral limit,
this model (for $G_S>1$) breaks chiral symmetry spontaneously
via an expectation value for the scalar quark-antiquark one-point function
(quark condensate).
We use here a cut-off in proper time as the regulator.
In the mean-field approximation \rcite{BBG1}
we can introduce the vacuum expectation value into the Lagrangian,
via an auxiliary field,
and then keep only the terms quadratic in
quark fields.
The ${\cal L}_{\rm NJL}^{\rm S,P,V,A}$ above
are then equivalent to a constituent chiral quark-mass term
of the type in \rcite{GM} of
the form $-M_Q (\bar{q}_L U^{\dagger} q_R + \bar{q}_R U q_L )$
\rcite{ERT}.

In this model, two-point functions are given by the general graph
depicted in Fig.  \tref{Fig2pt}.
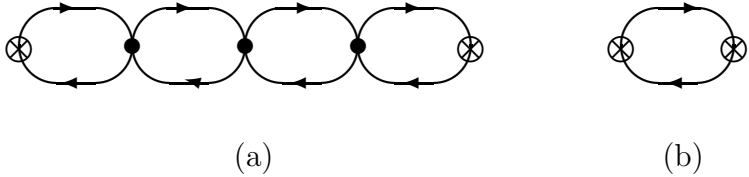
\begin{figure}
\begin{center}
%
%
%
\thicklines
\setlength{\unitlength}{1mm}
\begin{picture}(140.00,35.00)(0.,15.)
\put(97.50,35.00){\oval(15.00,10.00)}
\put(103.00,33.50){$\bigotimes$}
\put(17.50,35.00){\oval(15.00,10.00)}
\put(25.00,35.00){\circle*{2.00}}
\put(32.50,35.00){\oval(15.00,10.00)}
\put(40.00,35.00){\circle*{2.00}}
\put(47.50,35.00){\oval(15.00,10.00)}
\put(55.00,35.00){\circle*{2.00}}
\put(62.50,35.00){\oval(15.00,10.00)}
\put(08.00,33.50){$\bigotimes$}
\put(68.00,33.50){$\bigotimes$}
\put(88.00,33.50){$\bigotimes$}
\put(38.50,19.00){(a)}
\put(95.50,19.00){(b)}
\put(14.50,40.00){\vector(1,0){3.00}}
\put(29.50,40.00){\vector(1,0){3.50}}
\put(44.00,40.00){\vector(1,0){5.00}}
\put(60.50,40.00){\vector(1,0){3.00}}
\put(95.50,40.00){\vector(1,0){5.00}}
\put(99.00,30.00){\vector(-1,0){3.00}}
\put(64.00,30.00){\vector(-1,0){3.00}}
\put(49.50,30.00){\vector(-1,0){3.50}}
\put(34.00,30.00){\vector(-4,1){2.00}}
\put(18.00,30.00){\vector(-1,0){2.50}}
\end{picture}
\caption{The graphs contributing to the two point-functions
in the large $N_c$ limit.
a) The class of all strings of constituent quark loops.
The four-fermion vertices are either
${\cal L}^{\rm S,P}_{\rm ENJL}$ or
${\cal L}^{\rm V,A}_{\rm ENJL}$ in Eqs. \protect{\rref{ENJL2}} and
\protect{\rref{ENJL3}}.
The crosses at both ends are the insertion of the external sources.
b) The one-loop case.}
\rlabel{Fig2pt}
\end{center}
\end{figure}
How to sum these kind of strings of bubbles of constituent
quarks for two-point
functions, regularization independent relations obtained in this model,
the extension of the technique from two-point
to three-point functions including explicit chiral symmetry breaking,
discussion of the Weinberg Sum Rules in this model, how VMD
works in this model and more related phenomenological issues are treated
in Refs. \rcite{BRZ,BP2,BBR,prades2} and reviewed in \rcite{Physrep}.
The general conclusion is that within its limitations the
ENJL-type models do include a reasonable amount of the expected physics
from QCD, its symmetries,
their spontaneous breakdown and even some of its short distance information,
as for instance the one embodied in the Weinberg Sum Rules.
This is a very important
point and has been one more of the reasons why we have chosen this
model. Relations like the Weinberg Sum Rules are theorems of QCD and
should be reproduced by any reasonable candidate to describe
 the low-energy dynamics.
In fact, they are essential in the convergence of the hadronic contribution
to the electromagnetic $\pi^+ - \pi^0$ mass difference \rcite{Dass}.
These relations guarantee the good matching between the low-energy
behaviour and the high-energy one. Models to introduce
vector fields like the Hidden Gauge
Symmetry (HGS) do not always have this good intermediate behaviour.
This HGS model was used in ref. \rcite{Gerard}
to estimate vector meson contributions to $B_K$ in an $1/N_c$
expansion.

 The major drawback of the ENJL model is the lack of a confinement mechanism.
Although one can always introduce an {\em ad-hoc} confining
potential doing the job.
We will smear the consequences of this drawback
by working with internal and external
momenta always Euclidean with $|p^2|<$ $\Lambda_\chi^2$. In the
$1/N_c$ expansion, we will also only keep singlet color observables.

We will use the ENJL model as a model to fairly describe
in the large $1/N_c$  limit
the strong interactions between the lowest-lying mesons and,
if needed, external sources.  The model we are using is
a tree-level loop model with a explicit cut-off regularization
for one loop parts. This introduces the physical cut-off $\Lambda_\chi$.
What we mean by a tree level loop model is the following. A general set
of external fields is connected via a set of one-loop diagrams with three or
more legs (vertices) and sums over connections of these vertices by
a four-fermion
 interactions or a full chain like depicted in Fig. \tref{Fig2pt}.
These are also the contributions which are leading in $1/N_c$.
Going beyond the tree-level approximation is
going beyond the large
$N_c$-limit. It is at this level that the hadronic properties of this model
have been tested. Also to going beyond this one would have to include
other operators not
suppressed at the next-to-leading order in $1/N_c$
in the ENJL Lagrangian. At that level one also encounters the problem of
regularizing overlapping divergences in the model.
For the purpose we want to apply the model here, namely, for calculating
next-to-leading $1/N_c$ corrections to hadronic matrix elements of
four-quark operators to consider strings of bubbles is sufficient
for the non-factorizable part, see Sect. \tref{method}. As discussed there
the leading non-factorizable part is calculable by planar diagrams. This
corresponds to the tree-level loop calculation.

\section{The ENJL Calculation}
\rlabel{calculation}
\setcounter{equation}{0}

In this Section we will explain how the calculation was done
in the ENJL model.
In the large $N_c$ limit there is just one kind of diagrams that
can contribute. These are depicted in Fig. \tref{FigLONJL}.
\begin{figure}
\begin{center}
\leavevmode
\epsfxsize=12cm\epsfbox{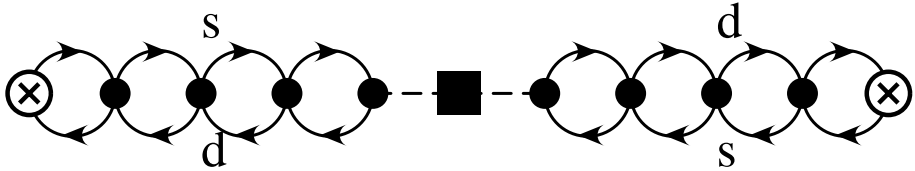}
\end{center}
\caption{The leading $1/N_c$ contribution to $\Pids(q^2)$
in the ENJL model.
Symbols as in Fig. \protect{\tref{FigQCD}}  except that a dot is now a
ENJL-vertex in Eq. \protect{\rref{ENJL1}}
and the full lines are constituent quark-lines.
The box is a $\Gamma_{\Delta S=2}$ operator insertion.
The flavour is mentioned next to the lines.}
\rlabel{FigLONJL}
\end{figure}
There the fermion lines
are constituent quarks with mass $M_d$ and $M_s$ \rcite{BP2}.
 We have the product of two
two-point functions, namely two mixed pseudoscalar--axial-vector
$\Pi_P^\mu(q)_{sd}$. These two-point functions are connected by the
exchange of a $X$-boson between the two left currents.
At leading order in the $1/N_c$ expansion
(i.e., at ${\cal O} (N_c)$ in this case),
each one of the two  two-point functions are  sum of
all the strings of bubbles or loops, i.e. one, two, $\cdots$,
$\infty$ .
This type of diagrams factorizes when the $X$-boson propagator is cut.
In terms of the two-point function\rcite{BRZ,BP2}
\be
\Pi_P^\mu(q)_{sd} \equiv i \int {\rm d}^4 x \, e^{i q \cdot x}
\langle 0 | T \left( A^\mu_{sd}(x) P_{ds}(0) \right) |0 \rangle \, ,
\ee
with $A^\mu_{sd}(x) \equiv \bar s (x) \gamma^\mu \gamma_5 d(x)$, and
$P_{ds}(x) \equiv \bar d (x) i \gamma_5 s(x)$. Then, at leading order
in $1/N_c$, we have
\be
\rlabel{NJLLO}
\Pids(q^2) = - \frac{1}{2} \Pi^\mu_P (-q)_{sd} \Pi_\mu^P (q)_{sd} \, .
\ee
The two-point function needed here was calculated in the full ENJL model,
in the chiral limit in Ref. \rcite{BRZ}
and in Ref. \rcite{BP2} for non-zero quark masses. In both
cases to all orders in the external momenta expansion. The result  in Eq.
\rref{NJLLO} is
equivalent to $B_K(\mu) = \hat B_K = 3/4$
once $\Pids (q^2) $ is reduced to the
matrix element in Eq. \rref{defbk}.

At next-to-leading order in the $1/N_c$ expansion, we have two general
kind of diagrams. The one depicted in Fig. \tref{FigNJL1} and the one in
Fig. \tref{FigNJL2}.
\begin{figure}
\epsfxsize=14cm 
\epsfbox{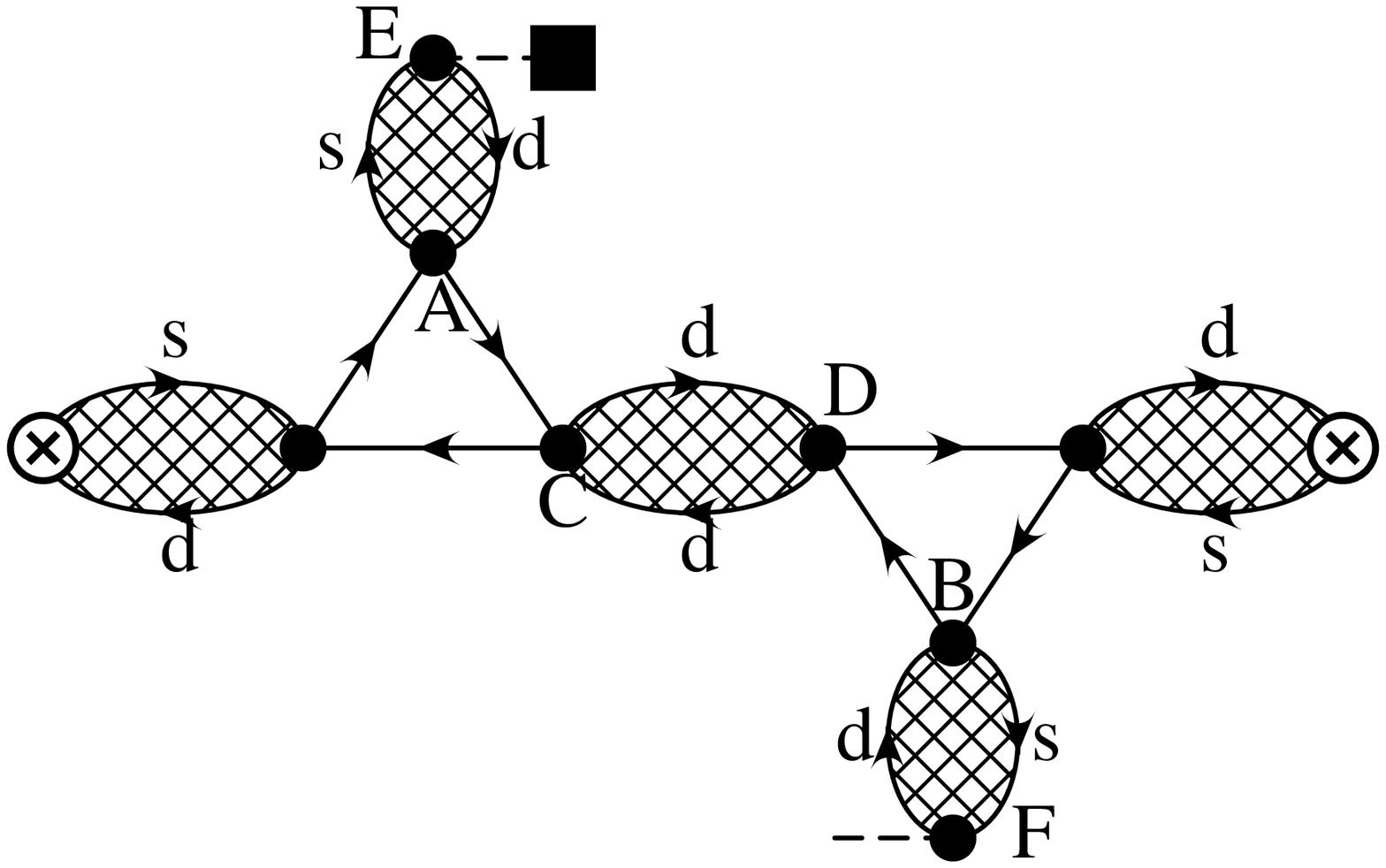}
\caption{The class of three-point diagrams. Symbols as in Fig.
\protect{\ref{FigLONJL}},
the hatched areas are a summation over sets of one-loop diagrams
as shown in Fig. \protect{\ref{Fig2pt}}.} There is also the same
type of diagram with the three-point functions tilted and the central
propagator $\bar d d$ changed to $\bar s s$.
\rlabel{FigNJL1}
\end{figure}
\begin{figure}
\epsfxsize=14cm 
\epsfbox{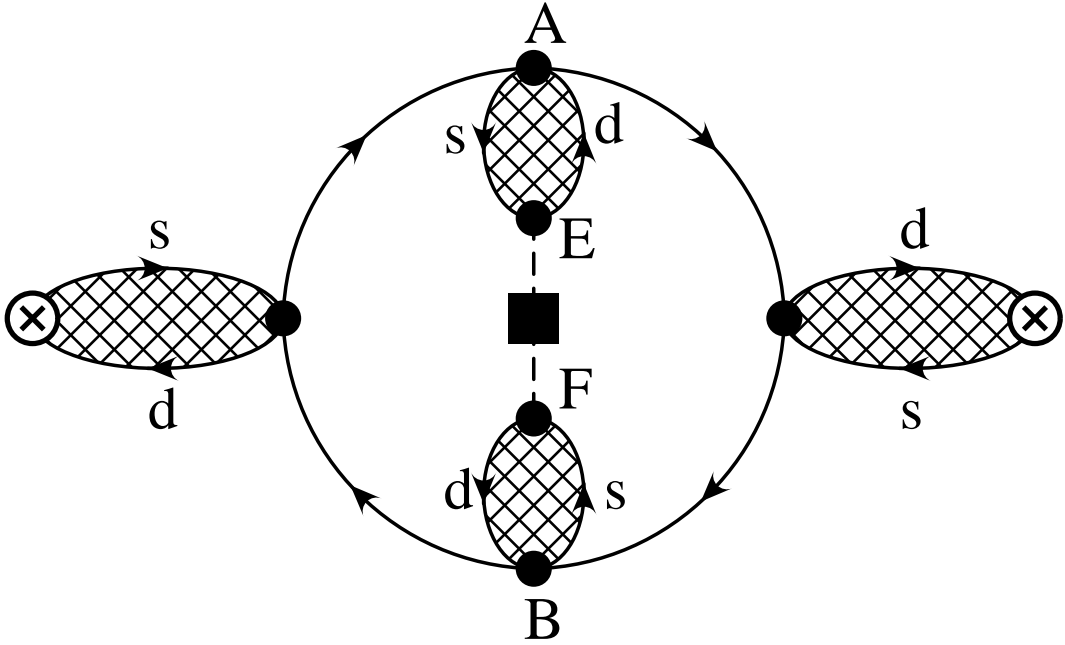}
\caption{The class of four-point diagrams, symbols as in Fig.
\protect{\ref{FigNJL1}}.}
\rlabel{FigNJL2}
\end{figure}
These are the two possibilities for the tree level loop diagrams and
correspond to Fig. \tref{FigQCD}b.
In both cases, when we cut the $X$-boson propagator, we have four
external legs. Two connected to the left currents and the other
two to the pseudoscalar sources. These are, then, tree-level constituent
quark loop
four-point  functions. The diagram in Fig. \tref{FigNJL1} can be seen as
the product of two one-loop three-point functions with a $P_{ds}$
 pseudoscalar leg, a left current $L_{sd}^\mu$ leg and the third leg
amputated. Then the two three-point functions are glued with a
propagator,
this can be $\bar s s$ or $\bar d d$ and with any kind of Dirac structure.
The other class contains a one-loop four-point function
dressed with two legs connecting the pseudoscalar sources $P_{ds}$
and another two connecting the left current sources $L^\mu_{sd}$.
Therefore, we have to calculate in the ENJL model, at leading order
in the $1/N_c$ expansion a generalized $\langle P^{ds} L_{sd}^\mu  P^{ds}
L^{sd}_\mu  \rangle$ four-point function. This implies the
actual calculation of many one-loop
four-point functions since the pseudoscalar
sources and left currents can mix with the other Dirac structures.
We have also to calculate all possible three- and two-point functions
made out of scalar, pseudoscalar, vector and axial-vector currents.
This, of course, is the major part of the work. For a more detailed
explanation of the kind of contributions we have to consider and
some examples of the diagrams in Figs. \tref{FigNJL1}
and \tref{FigNJL2}, see the appendix.

The calculation of these Green functions is done in the ENJL model.
The following discussion is for non-anomalous Green functions.
As emphasized in Sect. \tref{short},
this is a model of trees of bubbles where one bubble is
consistently regularized using a cut-off regulator. This cut-off
regulator has to preserve the QCD Ward identities. We do that by using
a proper-time regularization and imposing the QCD Ward identities,
i.e. adding the needed counterterms. Although,  proper-time
regularization breaks in principle these Ward identities one can
{\em always} add the counterterms that restore them.
Now, we would like to explain, in more detail, the way we have done the
regularization here.
After the Dirac algebra is done, we use relations like
$2 \, r \cdot q = ((q+r)^2-M^2)-q^2-(r^2-M^2)$ and $r^2 = (r^2-M^2)+M^2$
to reduce the number of propagators.
Then, the least divergent integrals
are just calculated by the standard way of rerouting momenta and reproducing
the dimensional regularization result when $\Lambda_\chi \to \infty$.
These integrals are the ones where all the Lorentz indices are saturated
by external momenta indices. This only introduces
${\cal O} \left(p^2/\Lambda_\chi^2 \right)$ ambiguities but they are of the
same order as is inherent in the applicability of
a non renormalizable model with just these two parameters,
$p^2$ and $\Lambda_\chi^2$.
Because of the results mentioned in the previous
Section we do not expect these
corrections to be very important.
In the rest of the integrals, some of the Lorentz indices are carried by
$g_{\mu \nu}$. These can in principle give rise to divergencies that
break the Ward identities. In fact the number of Ward identities in each
case is sufficient to determine these divergent parts. Thus,
we use all possible Ward identities to determine these kind of
integrals. In these way we ensure at the same time that our
Green functions fulfill the underlying QCD Ward identities.
In effect only a subset of the Ward identities is needed for this. The
remainder forms a check on the one-loop calculations.
This procedure is equivalent to using the heat-kernel expansion.
Applying
the prescription to reproduce the QCD flavour chiral anomaly consistently
given in Ref. \rcite{BP2}, we do not have, here, to add any counterterm
to the Feynman diagram calculation.

Once we have  consistently regularized two-, three- and four-point
functions we close the $X$-boson propagator by integrating in the loop
momentum $r$ up to the Euclidean cut-off $\mu$. The routing of the momenta
is the one depicted in the figures. As explained in
Sect. \tref{CHPTNc},  we reroute the
external momentum $q$ through the $X$ boson momentum $p_X = r+q$. The
presence of the cut-off in $r$ breaks the translational invariance on this
momentum and then the two-possible choices $p_X = r+q$ and $p_X = r-q$
give a numerical difference of the order of $q^2/\mu^2$.

\section{The Vector-Vector Toy Effective Lagrangian}
\rlabel{checks}
\setcounter{equation}{0}

Donoghue and Golowich \rcite{DG93}
suggested to look at a toy effective Lagrangian
that is a four-quark operator of the type vector current times vector
current.
The point is that the leading contribution to this type of effective
Lagrangian is calculable in terms of measurable quantities and it
may give some feeling on the low-energy behaviour of current times current
effective Lagrangians like the Standard Model one. However, the fact
that  the low-energy behaviour of this vector
current $\times$ vector current is controlled by leading order in
the $1/N_c$ expansion  measured spectral functions and thus
reliably calculable makes, at the same time,  its low-energy behaviour
quite different from the left current $\times$ left current
effective Lagrangian one. In this case  the low-energy behaviour
 is given by next-to-leading order at large $N_c$.

The toy $\Delta S=1$ hamiltonian proposed by \rcite{DG93}
is
\be \rlabel{toy}
{\cal H}^{\Delta S=1}_{VV} \equiv \frac{g_2^2}{8} \,
{\int } {\rm d}^4 \, x \, i D_{\mu \nu} (x, M_W) \, T \left( V^\mu_{1-i2}
(0) V^\nu_{4+i5} (x) \right) \, .
\ee
Where $g_2$ is the SU(2)$_L$ coupling and $G_F \equiv (g_2^2 \sqrt 2)
/(8 M_W^2)$. The currents are $V^\mu_{a+ib} (x) = \bar q (x)
\left[(\lambda_a + i \lambda_b)/2 \right] \gamma^\mu q (x)$ with
the following normalization tr$\left(
\lambda_a \lambda_b\right) = 2 \delta_{ab}$, and $\lambda_a$ are
the Gell-Mann
SU(3) flavour matrices. Then, they wrote down a sum rule relating,
in the chiral limit, the
amplitudes of the $K^- \to \pi^-$ and $\bar K^0 \to \pi^0$
transitions to measured vector $\rho_V(s)$ and axial-vector
$\rho_A(s)$ spectral functions.
The reduced amplitudes for these transitions
(see Ref. \rcite{DG93} for details) are proportional to
\be \rlabel{amplitude}
{\cal A} \equiv M_W^2 {\int^\infty_0} {\rm d} \, s \, s^2
\ln \left(\frac{s}{M_W^2} \right) \,
\frac{\rho_V(s) \rho_A(s)}{s-M_W^2+i\varepsilon} \, .
\ee

This sum rule is equivalent to the one obtained by Dass et al.
\rcite{Dass} for the
$\pi^+ - \pi^0$ electromagnetic mass difference,
\be
-\frac{32 \pi^2 F_0^2}{3 e^2} \,
\left( m_{\pi^+}^2-m_{\pi^0}^2 \right) \, = \,
 {\int^\infty_0} {\rm d} \, s \, s
\ln \left(\frac{s}{\Lambda^2} \right) \, \rho_V(s) \rho_A(s) \, .
\ee
Notice, that in the amplitude ${\cal A}$  the scale $M_W$ cannot be sent
to infinity. In the electromagnetic mass difference, the scale $\Lambda$
disappears due to the first and second Weinberg Sum Rules.

This vector $\times$ vector effective toy model in Eq. \rref{toy}
is related, in the soft pion limit,
to the vector (VV) part of diagram (d) in Fig. \tref{Figloops}
discussed in Sect. \tref{CHPTNc} through a SU(3) chiral rotation.
Diagram (d) is for the  $\bar K^0 \to K^0$ transition. It is then one of the
subdiagrams we have calculated for the $\Pids (q^2)$ in the ENJL
 model.  The amplitude in Eq. \rref{amplitude} can also be written
in the Landau gauge for the $W$ boson as
\be \rlabel{VV}
{\dis \int^{M_W^2}_0} {\rm d} s \, s^2 \left( \Pi_{V}^{(1)}(s)
- \Pi_{A}^{(1)}(s) \right) \, .
\ee
Where $\Pi_V^{(1)}(s)$ is defined as
\ba
&& i {\dis \int} {\rm d}^4 x \, e^{iqx} \langle
0 | T \left( V^\mu_a (0) V^\nu_b (x) \right) | 0 \rangle
\nonumber
\\&& \equiv \delta_{ab} \left[ \left( q^\mu q^\nu - q^2 g^{\mu \nu}
\right) \Pi_V^{(1)}(q^2) + q^\mu q^\nu \Pi_V^{(0)}(q^2) \right] \,
\ea
with $V^\mu_a (x) \equiv \bar q (x) \left[ \lambda^a/\sqrt 2 \right]
\gamma^\mu q(x)$. $\Pi_A^{(1)}(s)$ is defined analogously.

In the soft pion limit, the axial-vector (AA) part of diagram (d)
can be analogously related via a sum rule to the same
amplitude in Eq. \rref{VV} with the opposite sign.
Then at leading order (soft pion limit)
the left current $\times$ left current
effective Lagrangian does not contribute to this diagram
  and the result is ${\cal O}(1/N_c)$.

Following Ref. \rcite{BR91} we can split the integral above into
two regions separated by a cut-off $\mu$ $(0<\mu < M_W)$. For very small
$\mu$ one can use lowest order CHPT
\be \rlabel{CHPTVV}
\Pi_{V}^{(1)}(s) - \Pi_{A}^{(1)}(s)  \Longrightarrow
2 \frac{F_0^2}{s} - 8 L_{10}
\ee
where $L_{10}$ is one of the ${\cal O} (p^4)$ couplings
of the strong chiral Lagrangian \rcite{GasserLeutwyler}.
One can already see that the lowest part of the integral diverges
quartically. For scales $\mu$ very large one can use QCD \rcite{BBG2}
\be
\Pi_{V}^{(1)}(s) - \Pi_{A}^{(1)}(s) \Longrightarrow
 \frac{9 \pi}{2} \alpha_s(s)
\frac{\langle 0| \bar q q |0 \rangle ^2 (s)}{s^3} \, .
\ee
Then the upper of the integral diverges logarithmically.
One can use some kind of effective
low-energy model to improve the lowest order CHPT behaviour
in Eq. \rref{CHPTVV}
like it was done in Refs. \rcite{BRZ,BR91} for
the electromagnetic pion mass difference. There the ENJL model
and the constituent
quark model, respectively, were used obtaining a better matching
with QCD. However, the divergences here are quartic instead
of quadratic as there making it more difficult. In fact,
as noticed in Ref. \rcite{DG93} the largest contribution to the
sum rule above is for the range of energies between a few GeVs and 10 GeV.
Unfortunately,  in this region the constituent quark model or the ENJL
model cannot
help since they are intended for energies below or around
the symmetry scale breaking $\Lambda_\chi \simeq 1.2$ GeV.
The large mismatch between the
scale dependence of the VV (AA) part of diagram (d) at low
and high  energies reflects the large anomalous dimensions
of the vector-vector (axial-vector -- axial-vector)
four-quark effective operators. This points also in the direction of
getting large effective couplings for  vector - vector (axial-vector --
axial-vector) four-quark operators at low energy.

Nevertheless, we can still use the fact that the cancellation
between the VV and AA part of diagram (d) in Fig.
\tref{Figloops} is  exact in the soft
pion limit (i.e., $q^2= m_\pi^2 =0$) due to chiral symmetry alone.
This cancellation  will then be a check
of chiral symmetry for our ENJL calculation\footnote{This
cancellation between VV and AA parts is the same that
occurs in lattice  numerical simulations.
The cancellation seen there is thus a consequence of chiral
symmetry.}. The cancellation indeed happens to all the values of  $\mu$
calculated in this work with an accuracy better than 1 $\%$.
This is one more non-trivial check to add to the ones discussed
in previous Sections.

In our numerical results we can also separate out the result for the
VV (AA) part and relate the result obtained from our full
calculation for $q^2=m_K^2=0$ to the result obtained from
\be
\int_0^{\mu^2} {\rm d} s \, s^2
\left[ \Pi_V^{(1)}(s) - \Pi_A^{(1)}(s) \right]_{\rm ENJL}\ ,
\ee
in the chiral limit ($m_\pi^2 =0$).
This provides an explicit check of the PCAC relation in the ENJL model
and a very non-trivial check on our full calculation.

\section{Results}
\rlabel{numerics}
\setcounter{equation}{0}

In this Section we are going to discuss the results we get.
As in Ref. \rcite{BP1} we have studied three cases, namely, the chiral case
where we set  the quark masses to zero; the case with SU(3) explicit
symmetry breaking $m_s \neq m_d$ and the case with non-zero quark masses
but with $m_s = m_d$. This is the first time
that the chiral case is discussed separately and to all
orders in momenta.
This permits us to obtain directly the coupling $G_{27}$ introduced
in Sect. \tref{CHPT1}. The remaining cases are important to assess the size
of the corrections due to the presence of non-zero kaon and pion masses.
The case $m_d=m_s \neq 0$ is done because present lattice calculations
are done in that limit. In fact, it has been noticed recently \rcite{Gupta}
that quenched lattice data, which are used at present to predict
this type of matrix elements, cannot be used to extrapolate to
physical light quark masses. This leaves the error due to the use
of degenerate quarks and quenched QCD difficult to estimate when
extrapolating to the physical $B_K$.

All the values of input parameters used in the ENJL model are given
in Sect. \tref{short}. The physical parameters,
which are relevant for this calculation, obtained with these values
are $F_0=89$ MeV, $m_K=495$ MeV and  $m_\pi=135$ MeV.

The procedure we followed to analyze the numerical
results is also the one in Ref. \rcite{BP1}. We fit the ratio between the
$1/N_c$ corrections and the leading result for a fixed scale $\mu$
to $a/q^2 + b + c q^2 + d q^4$ which always give very good fits.
The $a$ term is the lowest one allowed by chiral symmetry. We also allowed
for one more term
here ($d$ term) to increase the accuracy of the extrapolation.
Here $a$, $b$, $c$ and $d$ are $\mu$ dependent
for the $B_K(\mu^2,q^2)$  form factor. Had we calculated just
to ${\cal O}(p^4)$,  these $a$, $b$ and $c$ terms would correspond
to the $A$, $B$ and $C$ terms introduced in Sect. \tref{CHPT1}.
Notice that at the order in $1/N_c$ we are working CHPT predicts that
the $q^2$ dependence of $\Pids (q^2)$ is analytic to all orders in CHPT, see
Subsect. \tref{CHPT1}.3. So we expect to fit our numbers with this
form.

The values of the $q^2$
external momenta used to make  the fit are in the Euclidean region and
below the constituent quark production threshold. As said
previously, this is done to
smear the possible consequences of using a non-confining model.
Once we have these coefficients
we extrapolate the $B_K$ form factor to  $q^2=m_K^2$
to obtain the physical $B_K$.
For calculating $\hat B_K$ we use $a_+=-2/9$ and
$\alpha_s^{(1)}(M_\tau)
= 0.33 \pm 0.03$ \rcite{Tonitau} which corresponds to
$\Lambda_{\overline {MS}}^{(3)}= \left( 215 \pm 40 \right)$ MeV
to one loop. When the two-loop ${\rm{\overline {MS}}}$
running
for $\alpha_s (\mu)$ is done, this value is
in agreement with the LEP value for $\alpha_s (M_Z)$\footnote{
If the lower value $\alpha_s^{(1)}(1 {\rm GeV})=
0.336 \pm 0.011$, recently obtained in Ref. \rcite{Voloshin}
from the $\Upsilon$ system, is used then $\Lambda_{\overline
{MS}}^{(3)} = \left(125 \pm 15 \right)$ MeV the values for $\hat
B_K$ are slightly  larger and less stability is obtained.}. Since
the main source of error in our calculation is of hadronic
origin we prefer to give also the running $B_K(\mu)$ parameter.
In addition, this information can be used with any more accurate
vale of $\alpha_s^{(1)}$ one can get in the future.
The errors in Table~\tref{table1} are only from the uncertainty in
$\alpha_s$.

\begin{table}
\begin{center}
\begin{tabular}{c|cc|ccc|cc}
$\mu$ (GeV) & $B_K^\chi(\mu)$ & $\hat B_K^\chi$ & $B_K^m(\mu)$ &
$B_K^a(\mu)$ &
 $\hat B_K^m$ & $B_K^{\rm eq}(\mu)$ & $\hat B_K^{\rm eq}$ \\ \hline
0.3  & 0.60  &0.51(6) &    0.75 & $-$0.11    &0.64(8) &0.74 &0.63(8) \\
0.4  & 0.49 &0.48(2) &    0.72 & $-$1.6     &0.70(3) &0.72 &0.70(3) \\
0.5  & 0.35  &0.36(2) &    0.67 & $-$4.1     &0.70(3) &0.66 &0.69(3) \\
0.6  & 0.18  &0.20(1) &    0.57 & $-$7.5     &0.62(2) &0.56 &0.61(2) \\
0.7  & $-$0.05 & $-$0.06 &  0.46 & $-$12.     &0.52(1) &0.42 &0.47(1) \\
0.8  & $<$0  & $--$ &   0.30 & $-$16.     &0.35(1) &0.23 &0.27(1)\\
0.9  & $<$0  & $--$ &   0.11 & $-$21.     &0.13(1) &0.02 &0.02 \\
1.1  & $<$0  & $--$ &   $<$0 & $--$       & $--$  & $<$0 & $--$
\end{tabular}
\end{center}
\caption{Results for $B_K$ and $\hat B_K$. The uncertainty due
to $\alpha_s(M_\tau) =0.33\pm0.03$ is indicated in brackets.}
\rlabel{table1}
\end{table}

Let us start discussing the chiral case.
The corrections we get for the chiral case are very large and
negative. Unfortunately not very much else can be said due to the
lack of stability. We can only give a range of values for
the value of $\hat B_K$. For the lower bound, it is clear that
below 0.3 GeV the QCD $\mu$ scale dependence is not trustable.
For the upper bound one should remain roughly below the two constituent
quark threshold to be safe of possible effects due to the
constituent quark production. Therefore we propose for the chiral
limit the following range
\be
0.30< \hat B_K^\chi <0.50 .
\ee
 This corresponds to a
very narrow window in energy [(0.3  $\sim$ 0.5) GeV]. As said in Sect.
\tref{CHPT1} for the chiral case ($m_s=m_d=0$) we are actually
determining the $G_{27}$ coupling. The values of $G_{27}$ corresponding
to the range above are
\be
0.40< G_{27} <0.67 .
\ee
Notice that this value is compatible with the one obtained
in Eq. \rref{G27B} from the $K^+ \to \pi^+ \pi^0$
$\Delta I =3/2$ decay rate.
Moving away from $q^2=0$ we obtain a value for the slope $C$
or for $D_2^r+D_3^r/4$.

Let us go to the equal mass case ($m_s = m_d$). The results for this case
are in the 7th and 8th columns. Here, we also get that the corrections
to the leading $1/N_c$ behaviour are negative.
The fits done are good and
are compatible with the absence of the $a$ term.
This was also the case in the chiral calculation of Sect. \tref{CHPT1}
and provides another numerical check on our results.
We observe also that
corrections due to quark masses are positive and tend to bring
the value of $\hat B_K$ to its large $N_c$ result. Again, we have not
a very good matching with the leading QCD logarithmic corrections.
Although, is somewhat better than the one we got for the chiral case.
Though there is a narrow maximum around $\mu \simeq$ (0.4 $\sim$ 0.5)
GeV, this is just an artifact produced by the perturbative running of
$\alpha_s (\mu)$
for values as low as  $\mu \sim$ 0.3 GeV. Thus,
we prefer also not to give a central value for $B_K$ which will be
misleading. Then, for the lower bound we have the same argument as for
the chiral case. For the upper bound
we can enlarge a bit the window with respect
to the chiral limit due to the presence of nonzero quark masses
that makes the stability  broader. Then we propose for this
equal mass case the following range
\be
0.55 < \hat B_K^{eq} <0.70 \, .
\ee
Notice that in the equal mass case there is no shift needed due to
the difference in octet or nonet chiral logarithms.
{}From the fit we can
determine one combination of $D_7$ and $D_4+D_3$, namely, the $b$ term.
{}From the same fit we also get a value for $C$
from the $c$ term. The difference between this value for $C$ and the
one obtained in the chiral limit indicates the size of the $p^6$
corrections.

Let us now comment on the more realistic case,
($m_s \neq m_d$). This is the one
in columns 4th, 5th and 6th. In the 5th column we present the
$B_K(\mu,q^2)$ form factor for $q^2=-0.001$ GeV$^2$ showing the
presence of the pole term discussed in Sects. \tref{CHPT1}
and \tref{CHPTNc};i.e. $a \neq 0$. For small values of $q^2$ this pole
is the overriding numerical contribution. Its contribution to $\hat B_K$
is of the right size as expected for a higher order CHPT contribution.
The value of this pole term allows us to determine the value of $D_1$.
  Again the matching with the
leading QCD logarithmic corrections is not very good although  better
than for the other two cases. Clearly, in this approach,
the presence of masses stabilizes the $\hat B_K$ parameter.
As happened for the $G_V=0$ case in Ref. \rcite{BP1}, but now
more pronounced, the $a$ term dominates the $B_K(\mu,q^2)$
form factor for small values of $q^2$
even though is just a small fraction when extrapolated to the physical
$B_K$. From columns 6th and 8th one can conclude that
having degenerate quark masses does not have much influence
on the final values obtained for $\hat B_K$. This is
something that can be used for present lattice data, remember
that they are still done in the degenerate quark limit. So, from this
more realistic case and with the complete ENJL,
improving previous similar determinations \rcite{BBG1,Gerard},
 we will give what we
consider the result for $\hat B_K$ in the large $N_c$ expansion approach.
We also
prefer not to give a central value for the same reasons as given in the
previous case. In fact, we believe that also from the results
in \rcite{BBG1,Gerard} one cannot infer a central value and it would be more
realistic there to give a range of values.
Therefore, we propose from the
results obtained in this work the following range of values
\be
0.55 < \hat B_K < 0.70 \, .
\ee
These numbers can also be used to determine a second combination of $D_7$ and
$D_4+D_3$. So we have a handle on all separate coefficients that contribute
to $B_K$.

Let us give now the explicit values for the $D_i$'s counterterms obtained
from this calculation. As said above they can be
obtained from the $a$, $b$ and $c$ terms of the
fits we have done for each value of the scale $\mu$. They should be
$\mu$-scale independent if the matching with the perturbative QCD running
was perfect. As explained before, we believe that a safe range of scales
to make predictions from this ENJL calculation is for $\mu$ in between
0.3 GeV and 0.6 GeV. Then for this range of $\mu$
and for the chiral logs scale $\nu = m_\rho \simeq 0.77$ GeV we get
\ba
-5.1 \cdot 10^{-2} < \left[ 16 \pi^2 F_0^2 \right] D_1^r <
- 3.4 \cdot 10^{-2} , \nonumber \\
-2.4 \cdot 10^{-3} < \left[ 16 \pi^2 F_0^2 \right] \left( D_2^r
+ \frac{D_3^r}{4} \right) < 1.9 \cdot 10^{-3} , \nonumber \\
0.3 < \left[ 16 \pi^2 F_0^2 \right] \left( D_4^r +
D_3^r - 4 \frac{L_5^r}{F_0^2} \right) < 0.6 , \nonumber \\
- 0.27 < \left[ 16 \pi^2 F_0^2 \right] \left( D_7^r
- 8 \frac{L_4^r}{F_0^2} \right) < - 0.24 .
\ea
Here we have used the $m_s \neq m_d$ data.
To disentangle the
$D_7$ and  $D_4 + D_3$ terms, we have also used
the $m_s = m_d$ data  and the value of $G_{27}$ obtained from the
chiral case data. Then these terms will have higher order
(${\cal O} (p^6)$) chiral
symmetry errors proportional to quark masses. In addition, we know
that the $D_2 + D_3/4$
counterterm can also be obtained from the chiral
limit ($m_s = m_d =0$) data, then we get
\ba
4 \cdot 10^{-2} < \left[ 16 \pi^2 F_0^2 \right] \left( D_2^r
+ \frac{D_3^r}{4} \right) < 7 \cdot\ 10^{-2} \, .
\ea
The difference from the previous determination is huge
(one order of magnitude). This again indicates that actually
explicit chiral symmetry breaking corrections  are very large.
All these determinations have, of course,  order $p^6$
chiral corrections since we are just using CHPT to order $p^4$
to get the $D_i$'s.

The chiral symmetry breaking due to the presence
of quark masses can then be quantified to be
\be \rlabel{chiralbreaknon}
\frac{B_K}{B_K^\chi} \simeq 1.60 \pm 0.35 \,
\ee
in nonet symmetry and
\be \rlabel{chiralbreakoct}
\frac{B_K}{B_K^\chi} \simeq 1.8 \pm 0.4 \,
\ee
in octet symmetry.
This chiral symmetry breaking is actually very large. However it is
of the same order as others found in the same system, for instance
we have
\be \rlabel{chiralbreak2}
\frac{F_K^2}{F_0^2} \simeq 1.8 \, ,
\ee
which also appears in the matrix element of the ${\cal O}_{\Delta S=2}
(x)$ operator in Eq. \rref{defbk}.
In fact our results point towards an understanding of the discrepancies
between previous calculations of the $B_K$ parameter and,
in particular, the small
values obtained in  Ref. \rcite{Donoghue1},
the QCD-Hadronic Duality estimate \rcite{AP1,prades1}
and the QCD-effective action approach \rcite{AP2}, although the error
bars in this last one are quite large.
The lattice results
\rcite{lattice}  and the next-to-leading $1/N_c$ result of \rcite{BBG1}
give larger values and are contained within the results of the
present work.
The QCD Sum Rules estimations \rcite{QCD2} have much larger error bars
due to the uncontrolled scale dependence in their calculation of
$B_K(\mu)$. Within errors this QCD Sum Rules estimation is compatible
with both groups of results above.
The effects due to explicit quark
masses are different in all these different calculations
and the uncertainty due to these effects could not be  estimated
in any of them.
The difference between the results for $B_K$ of these calculations
is of the size of the explicit chiral symmetry breaking effects
 we have obtained here.

In general, our results were obtained with a ninth
pseudo-Goldstone Boson. This
is the correct thing to do in a strict $1/N_c$ calculation. One needs to
extend the ENJL model to lift the $\eta'$ to a higher mass. In the present
work,  if we assume that, as in the strong-semileptonic sector,
the coefficients
are well determined by a leading $1/N_c$ calculation and that the main
effect of octet-nonet symmetry breaking is in the loop calculation,
we can then take the estimate given in Sect. \tref{CHPT1} for this effect,
\be
\hat B_K = \left(\hat B_K\right)_{\rm nonet} + 0.09\ .
\ee
To obtain our final range for the $B_K$ parameter we will also add
an educated guess of
the $1/N_c^2$ corrections of this next-to-leading in $1/N_c$
calculation. Then the range we propose for the chiral case is
\be
0.25 < \hat B_K^\chi < 0.55\
\ee
and for the real $m_s \neq m_d$ case
\be
0.60 < \hat B_K < 0.80 \, .
\ee

\section{Conclusions}
\rlabel{conclu}
\setcounter{equation}{0}

In the present work we have calculated the $B_K$ parameter defined
in Eq. \rref{defbk} in the $1/N_c$ expansion. We have essentially used
the technique in Ref. \rcite{BBG1} with more emphasis
in the low-energy contributions
and systematic uncertainties. We therefore extended their method to
Green functions rather than on-shell amplitudes. The use
of this Green function, $\Pids$ allowed us to study several issues involved
in the calculation of non-leptonic matrix elements.

We have performed a complete
CHPT calculation of the $\Delta S=2$ $\Pids (q^2)$ two-point
function in Eq. \rref{twopoint}, and therefore of $B_K$ to ${\cal O}
(p^4)$. From this calculation we conclude that there  is
an unknown  counterterm whose contribution to the $B_K$
parameter vanishes in the limit $m_s = m_d$.
This is a source of uncontrolled error for present
lattice simulations which are done in this limit and therefore
making the estimation of the error for the
extrapolation from this case to the real
case unreliable unless the exact $m_s \neq m_d$ is done.
In addition the present
quenched data has another source of uncontrolled error when
extrapolating to real quark masses \rcite{Gupta}.

 Then we, using an explicit cut-off $\mu$
for the fictitious $\Delta S=2$ $X$-boson introduced in Sect.
\tref{method}, have used lowest order CHPT first and the ENJL model
afterwards, to compute the low-energy hadronic contributions to
this two-point function at next-to-leading ${\cal O}(1/N_c)$.
We have then studied the type of information one can get from this
kind of calculations for the counterterms appearing in a pure CHPT
like the one in Sect. \tref{CHPT1}. We studied three cases, namely
the chiral case $m_s=m_d=0$, degenerate strange and down quarks
and the real $m_s \neq m_d$ case. The chiral limit
is not easy to obtain in other popular
non-perturbative methods like lattice simulations, QCD sum rules
or QCD-Hadronic Duality. However, it provides very interesting
information since, in this limit, the $1/N_c$ correction to
the $B_K$ parameter is correlated with the $1/N_c$ correction
to the octet $\Delta I =1 /2$ dominating $g_8$ coupling \rcite{PR95}.
We have shown that this two-point function in fact allows to determine
all free parameters in CHPT to order $p^4$ needed for a study of $B_K$.

In general we obtain somewhat
less stability, here, in the complete ENJL model
than in $G_V=0$ case studied in Ref. \rcite{BP1}. This  is just telling us
that, as said there, the model with $G_V=0$ had stability in a region
where the spin-1 four-fermion interactions are important and cannot be
neglected. The study done for $G_V=0$ was to all orders in CHPT and now
we have also the complete model also to all orders in CHPT. This situation
is not quite the same as the one
in Ref. \rcite{BBG1}. There, first  lowest order
CHPT was used and then vector mesons added in VMD model
to enlarge the range of energies where the calculation
was reliable. There, as expected then, more stability (enlarged
region) was obtained when vector mesons were added.
Although the results of our calculation have slightly
less stability than those
in Ref. \rcite{BBG1}, we have, as said in Sect. \tref{short},
a model for including the vector  and axial-vector mesons
that matches with the QCD high energy behaviour. Namely, our ENJL
model possesses the first  and second  Weinberg Sum Rules. Then we believe
that the results we get are more realistic than the ones in Ref.
\rcite{BBG1}.
This lack of good stability (matching) also tells us that,
unfortunately, this ENJL model fails to reproduce the QCD perturbative
scaling. We have forced the model to higher scales to see its behaviour
compared with perturbative QCD. The cause of the failure can be traced in
the treatment of the gluon propagator. We have taken it to be local below
some cut-off scale $\Lambda_\chi \simeq 1.2$ GeV. This seems a too
drastic reduction for energies above $\sim$ 0.6 GeV. A more sophisticated
model allowing for a momentum dependent gluon propagator might help.
Nevertheless, we believe that the qualitative conclusions
 regarding the $B_K$ parameter and the very conservative numerical
ranges we propose, and in particular, for $\hat B_K$
and explicit chiral symmetry breaking obtained here, are correct.

The chiral corrections due to quark masses
are large and positive.
As remarked in \rcite{prades1}, in the QCD-Hadronic Duality approach
there are operators of higher dimensions that contribute
to $B_K$. These were partially taken
into account by using intermediate hadronic states with the
physical masses and widths. To do better than that would require
to derive the effective QCD Lagrangian to all orders in the expansion
in quark masses which is clearly beyond reach now. We have instead
used an ENJL cut-off to model the low-energy hadronic interactions
with the advantage that the effective Lagrangian can be derived to
all orders in CHPT and therefore give an estimate of those higher
dimensional operators.
Our result in Eq. \rref{chiralbreakoct} gives some hint
of why the QCD-Hadronic Duality approach in Refs. \rcite{AP1,prades1}
gives  lower values than the lattice results or
the $1/N_c$ calculations here and in Refs. \rcite{BBG1,Gerard}.
 In fact the QCD-Hadronic Duality approach
 gives  roughly the ones we get for the chiral case, these also coincided
with the PCAC in the chiral limit determination of \rcite{Donoghue1}.

With the dynamical assumptions presented in Sect. \tref{method},
we have also seen that the results for the equal mass case and the
real case are not very different. This conclusion is very important
for the present lattice data since they cannot be extrapolated to
the physical masses for the reasons explained above.
They still contain the
error due to the use of quenched QCD in the simulations.

We have also included in our final result the shift due to the
nonet approximation inherent in the $1/N_c$ limit
as well as an educated guess for the $1/N_c^2$ corrections.
 Our final result for the $\hat B_K$ parameter is:
\be
0.60  <  \hat B_K < 0.80
\ee
Implications of this for CP-violation phenomenology can be found
in several references, e.g. \rcite{BH92}.

\section*{Acknowledgements}
We thank Eduardo de Rafael for discussions.
This work was partially supported by NorFA grant 93.15.078/00.
JP thanks the
Leon Rosenfeld foundation (K\o{}benhavns Universitet) for support and
CICYT(Spain) for partial support under Grant Nr. AEN93-0234.

\appendix
\def\theequation{A.\arabic{equation}}
\setcounter{equation}{0}
\section*{Explicit expressions}
\rlabel{appendix}

Here we give some explicit expressions and sketch how the
four- and three-point-like diagrams are calculated.
We will then give an example for one three-point-like diagram
and another for four-point-like diagrams
corresponding to the Figures  \tref{FigNJL1} and \tref{FigNJL2}.
The notation for the four-fermion couplings used here with respect to
the one in Section \tref{short} is
\ba
g_S &\equiv& \frac{4 \pi^2 G_S(\Lambda_\chi)}{N_c \Lambda_\chi^2}
\nonumber \\
g_V &\equiv &\frac{8 \pi^2 G_S(\Lambda_\chi)}{N_c \Lambda_\chi^2}  \, .
\ea

The $\Delta S=2$ two-point function $\Pids(q^2)$ we have to calculate
(see Sect. \tref{method}) involves as intermediate stage the calculation
of the generalized four-point functions
$\langle P_{ds} (0) L_{sd}^\mu (x)  P_{ds} (y)  L_\mu^{sd} (z)
\rangle$, with $2 L^\mu (x)= (V-A)^\mu(x)$ to all orders in
the external momenta $q$ flowing through the pseudoscalar current sources
and $p_X=r+q$ through the left current sources.
This generalized four-point function will afterwards be reduced
to a $\Delta S=2$ two-point function by integrating in the $X$-boson
momentum with an explicit cut-off $\mu$.
We have to calculate
for instance the generalized four-point function
$\langle P_{ds} (0) A_{sd}^\mu (x)   P_{ds} (y)  A_\mu^{sd} (z)
\rangle$ (also the one exchanging $A^\mu$ by $V^\mu$ sources).
For definiteness, let us give one example for each of the
two general type of contributions to this generalized
four-point function. They correspond to the Figures
in \tref{FigNJL1} and \tref{FigNJL2}.
In each of the four-fermion vertices any Dirac
structure conserving the strong interaction symmetries is allowed.

The three-point-functions like contributions
(Fig. \tref{FigNJL1}) to this
generalized four point function $\langle P_{ds}(0) A_{sd}^\mu (x)
P_{ds}(y) A_\mu^{sd}(z) \rangle$
consists, then, of two full three-point functions
(all orders in external momenta and quark masses) with
one pseudoscalar current source $P_{ds}$ and one axial-vector
current source $A_{sd}^\mu$  each.
They are obtained by gluing to the one-loop three-point functions
full two-point function legs (any permitted by the strong interaction
symmetries)
to obtain the full structure (see \rcite{BP2}).
Then the third leg of both full three-point functions
is removed and the two three-point functions
are pasted together with a propagator, i.e. a full two-point function
with flavour either $dd$ or $ss$ plus a pointlike coupling.
This propagator can have  any Dirac structure  compatible
with the strong interaction symmetries. The flavours are also
the ones corresponding to the generalized four-point obtained
after  conserving flavour in each four-fermion vertex.
As an example, one contribution of this type to the
generalized four-point function
with flavour structure as indicated in Fig. \tref{FigNJL1}, is
\ba
\langle P A P A  \rangle &\equiv& \cdots \nonumber \\
&-&
\frac{i}{2} \left[ 1 + g_S  \Pi_P(q)_{sd}\right]^2
\, g_V \,
{\dis \int^\mu_0} \frac{{\rm d}^4 r}{(2 \pi)^4} \,
\left[ g_{\mu \rho} - g_V \Pi^A_{\mu \rho} (-q-r) \right]_{ds}
g^{\mu \nu} \nonumber \\
&\times& \,  \left( {\overline \Pi}^{\rho b} _{PAV} \right)^{sdd}
(q+r,-r) \, \left[ g_{bc} - g_V \Pi^V_{bc} (-r) \right]_{dd}
\nonumber \\ &\times&
 \left( {\overline \Pi}^{\delta c} _{PAV} \right)^{sdd} (-q-r,r) \,
\left[ g_{\nu \delta} - g_V \Pi^A_{\nu \delta} (q+r) \right]_{ds}
\nonumber \\ &+& \cdots \, .
\ea
The two- and three-point functions here are defined with the notation
used in Ref. \rcite{BP2}. The non-barred $n$-point functions
correspond to the full functions (all orders in external momenta and
quark masses) and the barred ones to the one-loop expressions
\rcite{BRZ,BP2}.
There are more than 320 contributions like this one. In these
three-point-like function contributions
one also can have products of
two anomalous three-point functions, i.e.
three-point functions  which are proportional to a Levi-Civita
symbol.
To include them consistently we followed the prescription given in Ref.
\rcite{BP3} and in particular here we do not need to add any counterterm
to the naive Feynman diagram calculation.

The four-point like functions contribution to the generalized
four-point function consists, then,
in full four-point functions with the same flavour and Dirac structure
as the generalized four-point function
$\langle P_{ds}(0) L_{sd}^\mu (x) P_{ds}(y) L_\mu^{sd}(z)
\rangle$. Each  of these full-four functions
is constructed by gluing to the one-loop four-point
function two-pseudoscalar current sources $P_{ds}$ and
two left current source $L_\mu^{sd}$ with the
full two-point functions permitted
by the symmetries of the strong interactions that gives the required
structure. As an example, one  full four-point function contribution
to the generalized four-point function being considered
 and with the flavour structure corresponding to Fig. \tref{FigNJL2}
is
\ba
\langle P A  P A \rangle &\equiv&
\cdots \nonumber \\ &+&
\frac{i}{2} \left[ 1 + g_S  \Pi_P(q)_{sd}\right]^2
{\dis \int^\mu_0} \frac{{\rm d}^4 r}{(2 \pi)^4} \,
\left[ g_{\mu \rho} - g_V \Pi^A_{\mu \rho} (-q-r) \right]_{ds}
g^{\mu \nu} \nonumber \\
&\times& \,
\left( {\overline \Pi}_{PAPA}^{\rho \delta} \right)^{sdsd}(q+r,q,-q-r) \,
\left[ g_{\nu \delta} - g_V \Pi^A_{\nu \delta} (q+r) \right]_{ds}
\nonumber \\  &+&  \cdots \, .
\ea
Here, the four-point functions notation follows
up the notation of the two- and three-point function notation
explained before and introduced in \rcite{BRZ,BP2}
in an obvious manner. There are 16 contributions of this kind.

The one-loop $n$-point functions
are regularized using a proper
time cut-off that introduces the scale $\Lambda_\chi$, see
Refs. \rcite{BRZ,BP2} for details. Then,
these two examples give idea of the kind of calculations one
has to perform.
We have performed several checks on our calculation.
All of these checks of our ENJL
model calculation, namely
Ward identities, various comparisons with lowest order CHPT
calculations,\ldots,  have been passed successfully.
see Sects. \tref{CHPTNc}, \tref{calculation} and  \tref{checks}.

\end{document}